\documentclass{aa}  

\usepackage{graphicx}
\usepackage{bm}
\usepackage{txfonts}
\usepackage{float}
\usepackage{natbib}
\bibpunct{(}{)}{;}{a}{}{,} % to follow the A&A style

\begin{document} 

   \title{A direct and robust method to observationally constrain the halo mass function via the submillimeter magnification bias: \\Proof of concept}
   \titlerunning{Halo Mass Function measurement with Magnification Bias}
   \authorrunning{Cueli M. M. et al.}
   %\title{Halo Mass Function with Submillimiter Galaxy Magnification Bias}

   %\subtitle{I. Overviewing the $\kappa$-mechanism}

\author{Cueli M. M.\inst{1,2}, Bonavera L.\inst{1,2}, Gonz{\'a}lez-Nuevo J.\inst{1,2}, Lapi A.\inst{3,4}}

  \institute{$^1$Departamento de Fisica, Universidad de Oviedo, C. Federico Garcia Lorca 18, 33007 Oviedo, Spain\\
             $^2$Instituto Universitario de Ciencias y Tecnologías Espaciales de Asturias (ICTEA), C. Independencia 13, 33004 Oviedo, Spain\\
             $^3$International School for Advanced Studies (SISSA), via Bonomea 265, I-34136 Trieste, Italy\\
             $^4$Institute for Fundamental Physics of the Universe (IFPU), Via Beirut 2, I-34014 Trieste, Italy}

    \date{Received xxx, xxxx; accepted xxx, xxxx}

% \abstract{}{}{}{}{} 
% 5 {} token are mandatory
 
   \abstract
   {}
   {The main purpose of this work is to provide a proof-of-concept method to derive tabulated observational constraints on the halo mass function (HMF) by studying the magnification bias effect on high-redshift submillimeter galaxies. Under the assumption of universality, we parametrize the HMF according to two traditional models, namely the Sheth and Tormen (ST) and Tinker fits, derive posterior distributions for their parameters, and assess their performance in explaining the measured data within the $\Lambda$ cold dark matter ($\Lambda$CDM) model. We also study the potential influence of the halo occupation distribution (HOD) parameters in this analysis and discuss two aspects regarding the HMF parametrization, namely its normalization and the possibility of allowing negative values for the parameters.}
   {We measure the cross-correlation function between a foreground sample of GAMA galaxies with spectroscopic redshifts in the range $0.2<z<0.8$ and a background sample of H-ATLAS galaxies with photometric redshifts in the range $1.2<z<4.0$ and carry out a Markov chain Monte Carlo algorithm in the context of Bayesian inference to check this observable against its mathematical prediction within the halo model formalism, which depends on both the HOD and HMF parameters.}
   {Under the assumption that all HMF parameters are positive, the ST fit only seems to fully explain the measurements by forcing the mean number of satellite galaxies in a halo to increase substantially from its prior mean value. The Tinker fit, on the other hand, provides a robust description of the data without relevant changes in the HOD parameters, but with some dependence on the prior range of two of its parameters. When the normalization condition for the HMF is dropped and we allow negative values of the $p_1$ parameter in the ST fit, all the involved parameters are better determined, unlike the previous models, thus deriving the most general HMF constraints. While all the aforementioned cases are in agreement with the traditional fits within the uncertainties, the last one hints at a slightly higher number of halos at intermediate and high masses, raising the important point of the allowed parameter range.}
   {}

   \maketitle
%-----------------
%----------------
%1. INTRODUCTION
%----------------
%-----------------
\section{Introduction}

Within the $\Lambda$ cold dark matter ($\Lambda$CDM) model, the hierarchical growth of dark matter perturbations in the early Universe is an essential assumption needed to account for galaxy formation. Due to its high temperature, baryonic matter could not have formed gravitationally self-bound objects so early had they not been subject to gravitational interactions of some other nature that could overcome thermal energy. The very early freeze-out of dark matter allowed it to start clustering long before big bang nucleosynthesis could take place, providing the necessary potential wells for baryons to fall into. As a consequence, the relevance of dark matter halos for the probing of large-scale structure is unquestionable and has motivated the search for a quantitative understanding of their mass distribution.
\par
The first attempt at estimating this quantity dates back over 40 years. The Press-Schechter formalism \citep{ps} provided an analytic form for the halo mass function (HMF) based on spherical collapse and initial Gaussian fluctuations which laid the groundwork for ever-increasing efforts to determine this quantity as accurately as possible. An alternative derivation of the Press-Schechter HMF was carried out by Bond et al. within the so-called excursion set approach \citep{bond}.
\par 
Up until the end of the 1990s, the Press-Schechter mass function agreed reasonably well with most numerical simulations. However, as their resolutions improved, important deviations began to manifest themselves for halos below and above the so-called characteristic mass scale $M^{*}$, overestimating the former and underestimating the latter \citep{st1999}. The dynamics of ellipsoidal collapse were successfully applied to the excursion set formalism \citep{sheth01} and resulted in the widely used Sheth and Tormen (ST) parametrization of the HMF, which provides a very good fit when tested against N-body simulations. For instance, using high-resolution simulations for different cosmologies, \cite{J2001} showed that the HMF is fairly well described by the ST fit in the mass range from galaxies to clusters and from redshift 0 to 5. They suggested an alternative fit that provides some improvement at the high-mass tail but cannot be extended beyond said mass range. Moreover, they showed that the mass function could be expressed in a universal form when appropriately rescaled, meaning that the same analytical form and parameters could be used for different redshifts and cosmologies.
\par
Subsequently, a variety of fits to the HMF based on N-body simulations for different mass and redshift ranges were proposed, some of them confirming universality within a few percent \citep{reed03,reed07,warren06}, others quantifying small departures from it \citep{tinker08, crocce10,courtin11,watson13}. The question of universality is indeed a lenghty matter to discuss. However, as shown by \cite{despali16}, departures from universality could be associated with the way halos are defined \citep[see][for a summary of different halo finding methodology in simulations]{knebe13}. 

In essence, two common ways to obtain a halo catalog from an N-body simulation are friends-of-friends (FoF) algorithms \citep{davis85} and spherical overdensity (SO) algorithms \citep{lacey94}. Since there is not a universal definition of a dark matter halo, both methods have benefits and drawbacks and departures from a universal behavior have been found for the two kinds of algorithms. However, \cite{despali16} showed that, if SO-defined halos are defined using the virial overdensity (as opposed to other common criteria) and the mass function is expressed in terms of a parameter accounting for it, universality can then be retrieved to within a few percent. Their results were in agreement with those of \cite{courtin11}, who concluded that deviations from universality could be accounted for if one incorporates the redshift and cosmology dependence of the linear collapse threshold and the virialization overdensity.

Moreover, physical processes associated with baryons such as radiative cooling, star formation or feedback from supernovae and active galactic nuclei (AGN) have been shown to produce non-negligible modifications in the HMF, the effects being however sensitive to the modeling of the baryonic component. Indeed, \cite{cui12b} compared a dark-matter-only simulation with hydrodynamical counterparts without feedback from AGN, obtaining an increase in the number density of high-mass objects. However, the addition of AGN feedback by \cite{cui14} causes the opposite effect, a trend that has been confirmed using higher-resolution simulations, where a general decrease in the HMF is reported, more noticeable at low masses and redshifts \citep{sawala13,bocquet15,castro20}. Lastly, there could be physics beyond the Standard Model with a non-negligible effect on structure formation. Indeed, some authors have studied the inclusion of massive neutrinos \citep{costanzi13} or the effect of an interaction between dark energy and cold dark matter \citep{cui12a}. An effort toward observational constraints on the HMF could therefore provide some insight into these questions in addition to a validation of the results from N-body simulations. 
\par
Although some recent studies have provided observational methods to determine the HMF \citep{castro16,sonnenfeld19,li19}, all of them suffer from the uncertainties that arise when observational properties of cosmic structures are linked to the underlying halo mass. Our goal is not to assign halo masses to galaxies (or any of their observational properties) and empirically construct the HMF from there. In other words, we do not make use of a mass-richness relation, nor do we aim at obtaining one. We propose instead the use of an observable that, given its direct dependence on the halo mass and clustering of the foreground lenses, provides a robust measurement of the HMF. This physical quantity is the foreground-background galaxy angular cross-correlation function, together with background samples of submillimeter galaxies, which we argue to be promising candidates for cosmological analysis through the magnification bias effect \citep{gonzaleznuevo17,bonavera19, bonavera20, GON20}. We term this observable the submillimeter galaxy magnification bias.
\par
The aim of this paper is therefore to study two different HMF universal fits (namely the ST and Tinker models) with the aim of constraining their parameters and providing bounds to the HMF itself. This will be done by computing the angular cross-correlation function between two source samples with nonoverlapping redshift distributions and fitting the result through a Markov chain Monte Carlo (MCMC) algorithm to its theoretical prediction within the halo model formalism. Although the constrained HMF is in principle only representative of the galaxies producing the lensing effect, the comparison of the auto- and cross-correlation results by \cite{bonavera20} shows that the lens properties are indistinguishable from the galaxy parent population.
\par
The paper has been structured as follows. Section 2 provides a theoretical description of the physical situation. The usual formalism describing the HMF is presented, as well as a description of the chosen parametrizations. We also discuss the halo model prediction for our observable, the foreground-background angular cross-correlation function. Section 3 describes the methodology followed in our work process. We describe in detail the background and foreground galaxy samples as well as the cross-correlation measurement method. The MCMC algorithm used to fit the data to the model is presented, as well as the different runs we perform. Section 4 provides a discussion of the main results we obtained for the ST and Tinker fits and Section 5 details some further studies on the non-normalization of the HMF and the non-positivity of its parameters. The values for the $z=0$ HMF at certain masses are also given for the cases addressed in this work. The summary and our conclusions are given in Section 6, along with some ideas for future prospects.

%-------------------------
%-------------------------
%2. THEORETICAL BASIS
%--------------------------
%-------------------------
\section{Theoretical basis}

%2.1. The halo mass function
\subsection{The halo mass function}
\noindent
The common strategy when studying the statistical properties of mass fluctuations is to consider the overdensity field linearly extrapolated to the present, $\delta_0(\bm{x})$, and smooth it with a filter of scale $R$, that is, 
\begin{equation*}
    \delta_0^{R}(\bm{x})\equiv\int d^3\bm{x'}\delta_0(\bm{x'})W(\bm{x}+\bm{x'};R)=\int d^3\bm{k}\hat{W}(\bm{k}R)\delta_{0,\bm{k}}e^{i\bm{k}\cdot\bm{x}},
\end{equation*}
where $\hat{W}(kR)$ is the Fourier transform of the filter function $W(\bm{x};R)$, which, for the case of a top-hat in real space is given by
\begin{equation*}
    \hat{W}(kR)=\frac{3[\sin{kR}-kR\cos{kR}]}{(kR)^3}.
\end{equation*}
If we associate a mass $M$ with a comoving scale $R$ via
\begin{equation*}
    M=\frac{4}{3}\pi R^3\rho_0,
\end{equation*}
where $\rho_0$ is the mean matter density of the Universe at present time, we can interchangeably characterize a filter by its mass or length scale. The mass variance of the filtered linear overdensity field is thus
\begin{equation*}
    \sigma^2(M)\equiv\langle [\delta_0^{R}(\bm{x})]^2\rangle=\frac{1}{2\pi^2}\int_0^{\infty}k^2P(k)\hat{W}^2(kR)dk,
\end{equation*}
where $P(k)$ is the linear matter power spectrum at redshift $z=0$.
\par
Although its physical definition is clear, the mathematical parametrization of the HMF varies widely in the literature, so care must be taken when comparing results and different models. The (differential) HMF $n(M,z)$ is the comoving number density of halos at a given redshift per unit mass, that is,
\begin{equation*}
    n(M,z)dM
\end{equation*}
is the comoving number density of halos of mass in the range $[M,M+dM]$ at redshift $z$.
\par
One common way to parametrize it, which arises naturally from the excurstion set formalism, is
\begin{equation}
    n(M,z)=\frac{\rho_0}{M^2}f(\nu,z)\Big|\frac{\partial\ln\nu(M,z)}{\partial\ln M}\Big|,\label{n_mass}
\end{equation}
where $\rho_0$ is the comoving mean matter density of the Universe and\footnote{It should be noted that other authors define $\nu(M,z)$ without the square.}
\begin{equation*}
    \nu(M,z)\equiv\bigg[ \frac{\hat{\delta}_{\text{c}}(z)}{\sigma(M,z)}\bigg]^2,
\end{equation*}
with $\sigma^2(M,z)\equiv D^2(z)\sigma^2(M)$, where $D(z)$ is the linear growth factor for a $\Lambda$CDM universe, and $\hat{\delta}_{\text{c}}(z)$ is the linear critical overdensity at redshift $z$ for a region to collapse into a halo at that same redshift according to the spherical collapse model\footnote{The redshift dependence of $\hat{\delta}_c(z)$ is weak and usually neglected, that is, $\hat{\delta}_c(z)\approx 1.686$ for all $z$. However, we have taken it into account via the fitting function from \cite{kitayama96}.}. It is clear that $\nu$ depends on redshift and cosmology. However, if the function $f(\nu,z)$ is the same for all redshifts and cosmologies, that is, if $f(\nu,z)\equiv f(\nu)$ for all cosmologies, the mass function is said to be universal.
\par
For instance, the ST and Tinker $z=0$ models for the mass function are expressed in this parametrization as
\begin{align}
    f_{\text{ST}}(\nu)&=A_S\sqrt{\frac{a_S\nu}{2\pi}}\bigg[1+\bigg(\frac{1}{a_S\nu}\bigg)^{p_S}\bigg]e^{-a_S\nu/2}\label{stfit}\\
    f_{T}(\nu,z=0)&=A_T\bigg[1+\bigg(B_T\sqrt{\nu}\bigg)^{p_T}\bigg]e^{-C_T\nu},\label{tinkerfit}
\end{align}
where 
\begin{align*}
    A_S=0.322\quad\quad a_S=0.707\quad\quad p_S=0.3,
\end{align*}
and
\begin{align*}
    A_T=0.093\quad B_T=\frac{2.57}{\hat{\delta}(0)}\quad
    C_T=\frac{1.19}{\hat{\delta}^2(0)}\quad p_T=1.47.
\end{align*}
%and
%\begin{align*}
%    A_B(z)&=0.333(1+z)^{-0.11}\quad a_B(z)=0.788(1+z)^{-0.01}\\
%    p_B(z)&=0.807\quad q_B(z)=1.795
%\end{align*}
It should however be noted that some authors parametrize the HMF solely in terms of $\sigma(M,z)$, and care should be taken when relating the parameters from each definition.
\par
Furthermore, Sheth \& Tormen \citep{st1999} imposed a normalization condition, which in our parametrization reads
\begin{equation}
    \int_0^{\infty}\frac{f(\nu)}{M}\bigg|\frac{\partial\log{\nu}}{\partial\log{M}}\bigg|\,dM=1\label{normalization}
\end{equation}
and which accounts for the assumption that all mass is bound up in halos. As a consequence, their numerical fitting only dealt with two parameters ($a_S$ and $p_S$), since the normalization parameter ($A_S$) was fixed by \eqref{normalization}, yielding
\begin{equation*}
    A_S(p)=\Big[1+\frac{2^{-p}}{\sqrt{\pi}}\Gamma(1/2-p)\Big]^{-1}
\end{equation*}
and the condition $p<1/2$. Most authors have fit these models or their own to numerical simulations without imposing condition \eqref{normalization}, thus having an extra parameter. Although we find it more coherent with our halo model description to employ it in this work, we will only do so for the ST fit, since the Tinker fit as shown in equation \eqref{tinkerfit} cannot be normalized in this manner\footnote{The results one would obtain using the lesser-known normalizable Tinker fit are qualitatively similar.}.
\par
Given the fact that they are the most important and most widely used models, our analysis focuses on these two universal fits for the HMF: namely, a two-parameter ST fit,
    \begin{equation}
        f_{1}(\nu;a_1,p_1)=A_1(p_1)\sqrt{\frac{a_1\nu}{2\pi}}\bigg[1+\bigg(\frac{1}{a_1\nu}\bigg)^{p_1}\bigg]e^{-a_1\nu/2}\label{stlikefit},
    \end{equation}
and a four-parameter Tinker-like fit,
    \begin{equation}
        f_{2}(\nu;A_2,B_2,C_2,p_2)=A_2\bigg[1+\bigg(B_2\sqrt{\nu}\bigg)^{p_2}\bigg]e^{-C_2\nu}\label{tinkerlikefit}.
    \end{equation}

%2.2. The foreground-background angular cross-correlation
\subsection{The foreground-background angular cross-correlation}
\noindent 
The standard halo model considers that the matter density field at a point in space can be thought of as a sum over the density profiles of halos. In this context, the galaxy-dark matter cross-power spectrum can be parametrized by
\begin{equation*}
    P_{\text{g-dm}}(k,z)=P_{\text{g-dm}}^{\text{1h}}(k,z)+P_{\text{g-dm}}^{\text{2h}}(k,z),
\end{equation*}
where $P_{\text{g-dm}}^{\text{1h}}$ is the so-called 1-halo term, accounting for contributions within the same halo and $P_{\text{g-dm}}^{\text{2h}}$ is the so-called 2-halo term, accounting for contributions among different halos.
\par
These two quantities can be further expressed \citep{cooray02} as
\begin{align}
    P_{\text{g-dm}}^{\text{1h}}(k,z)&=\int_0^{\infty} dM\,M\frac{n(M,z)}{\bar{\rho}(z)}\frac{\langle N_{g}\rangle_M}{\bar{n}_g(z)}|u_{\text{dm}}(k,z|M)||u_{\text{g}}(k,z|M)|^{p-1}\label{crosspower1h}\\
    P_{\text{g-dm}}^{\text{2h}}(k,z)&=P(k,z)\Big[\int_0^{\infty}dM\,M\frac{n(M,z)}{\bar{\rho}(z)}b_1(M,z)u_{\text{dm}}(k,z|M)\Big]\,\cdot\nonumber\\
    &\quad\quad\quad\cdot\Big[\int_0^{\infty}dM\,n(M,z)b_1(M,z)\frac{\langle N_g \rangle_M}{\bar{n}_g(z)}u_g(k,z|M)\Big]\label{crosspower2h},
\end{align}
where $b_1(M,z)$ is the linear deterministic halo bias, $\bar{\rho}(z)$ is the mean matter density of the Universe, $\bar{n}_g(z)$ is the mean number density of galaxies, $\langle N\rangle_M$ is the mean number of galaxies in a halo of mass $M$ and $u(k,z|M)$ is the normalized Fourier transform of the matter distribution (be it dark matter or galaxies). Some comments should be made concerning \eqref{crosspower1h} and \eqref{crosspower2h}. Firstly, it is a reasonable approximation \citep{shethdiaferio01} to set the Fourier transform of the galaxy distribution to that of dark matter. Secondly, the mean number of galaxies within a halo of mass $M$ is split into a contribution from central galaxies and a contribution from satellite galaxies, parametrizing it in terms of the halo occupation distribution (HOD) parameters $\alpha$, $M_{\text{min}}$ and $M_1$, following \cite{zehavi05} and \cite{zheng05}. Lastly, the exponent $p$ should be set to $1$ for central galaxies and to $2$ for satellites \citep{cooray02}. More detailed information concerning the computation of all these quantities can be found in Appendix A.
\par
This cross-correlation between galaxies and dark matter can be probed via the weak lensing tangential shear-galaxy correlation \citep{bartelmann01} or via the foreground-background source correlation function. This work exploits the latter method, which is based on the fact that foreground sources trace the mass density field affecting the number counts of background sources.
\par
Indeed, in the presence of lensing, number counts observed in direction $\bm{\theta}$ and exceeding a flux $S$ are modified according to
\citep{bartelmann01}
\begin{equation*}
    N_S(\bm{\theta})=N^0_S\mu^{\beta-1}\,(\bm{\theta}),
\end{equation*}
where $N^0_s$ denotes the intrinsic source number counts exceeding flux $S$, $\beta$ is their logarithmic slope and $\mu(\bm{\theta})$ is the magnification factor in direction $\bm{\theta}$. In the weak-lensing limit, $\mu(\bm{\theta})\approx 1+2\kappa(\bm{\theta})$, where $\kappa(\bm{\theta})$ is the convergence. As a consequence, the fluctuations in the background number counts, which are due to magnification bias, can be written as 
\begin{equation*}
    \delta N_b(\bm{\theta})\equiv \frac{N_b(\bm{\theta})}{\bar{N}_b}-1=\mu^{\beta-1}(\bm{\theta})-1\approx 2(\beta-1)\kappa(\bm{\theta}).
\end{equation*}
Concerning the foreground sources, since they are supposed to trace the density field, the fluctuations in their number counts are due to pure clustering, that is,
\begin{equation*}
    \delta N_f(\bm{\theta})\equiv \int_0^{\chi_H}d\chi\, g_f(\chi)\,\delta_g(\bm{\theta},\chi),
\end{equation*}
where $\chi_H$ denotes the comoving radial distance to the horizon and $g_f(\chi)$ is the radial distribution of foreground sources.
\par
The angular cross-correlation between the foreground and background sources is then given by \citep{cooray02}
\begin{align}
    w_{fb}(\bm{\theta})&\equiv\langle\delta N_f(\bm{\varphi})\,\delta N_b(\bm{\varphi}+\bm{\theta})\rangle=\nonumber\\&=2(\beta-1)\int_0^{\chi_H}d \chi g_f(\chi)\hat{W}^{\text{lens}}(\chi)\,\cdot\nonumber\\
    &\cdot\int_0^{\infty}dk\frac{k}{2\pi}P_{\text{g-dm}}(k,z)J_0(kd_A\theta)\label{crosscorrcom},
\end{align}
where $\theta=|\bm{\theta}|$,  $d_A(\chi)$ is the comoving angular diameter distance, $J_0$ is the zeroth-order Bessel function of the first kind and
\begin{equation*}
    \hat{W}^{\text{lens}}(\chi)=\frac{3}{2}\frac{H_0^2}{c^2}a^2(\chi)E^2(\chi)\int_{\chi}^{\chi_H}d\chi'\frac{d_A(\chi)d_A(\chi'-\chi)}{d_A(\chi')}g_b(\chi').
\end{equation*}
In terms of redshift, \eqref{crosscorrcom} becomes
\begin{align}
    w_{fb}(\bm{\theta})&=2(\beta-1)\int_0^{\infty}\frac{dz}{\chi^2(z)}\,n_f(z)W^{\text{lens}}(z)\,\cdot\nonumber\\
    &\cdot\int_0^{\infty}dl \frac{l}{2\pi}P_{\text{g-dm}}(l/\chi(z),z)J_0(l\theta)\label{crosscorrang},
\end{align}
where we have defined $l\equiv kd_A(z)$,
\begin{equation*}
    W^{\text{lens}}(z)=\frac{3}{2}\frac{H_0^2}{c^2}\bigg[\frac{E(z)}{1+z}\bigg]^2\int_z^{\infty}dz'\frac{\chi(z)\chi(z'-z)}{\chi(z')}n_b(z'),
\end{equation*}
and $n_b(z)$ ($n_f(z)$) is the unit-normalized redshift distribution of the background (foreground) sources. $\beta$ is the logarithmic slope of the background source number counts and it is commonly fixed to 3 for submillimeter galaxies \citep{LAP11,LAP12,CAI13,BIA15,BIA16,gonzaleznuevo17,bonavera19}. In  this model, $\beta$ provides a general normalization whose possible changes are almost fully balanced by variations of $M_{min}$ (e.g., a $\approx$15\% increase in $\beta$ corresponds to a $\log{M_{min}}$ reduction of $\approx$1\%).
\par 
The foreground-background angular cross-correlation function \eqref{crosscorrang} clearly depends on the HMF parameters, given its prior dependence on $P_{\text{g-dm}}$. Therefore, we used this observable to constrain such parameters. Moreover, aside from the HMF parameters, the cross-correlation function depends on both the cosmology and the HOD parameters. Throughout our analysis, which assumes universality of the mass function, we keep the cosmology fixed to Planck's \citep{planck18VI} but aim to discuss the role of the HOD parameters by also including them in the MCMC analysis in some cases, as will be described in Section 3.3. 
%----------------
%----------------
%3.WORK METHODOLOGY
%----------------
%----------------
\section{Work methodology}

%3.1. Data
\subsection{Data}
\noindent
The background and foreground samples have been selected as described in detail in \cite{gonzaleznuevo17} and \cite{bonavera19}. 
The foreground sources consist of a sample of the GAMA II \citep{DRI11,BAL10,BAL14,LIS15} spectroscopic survey, with $0.2 < z < 0.8$. It is made up of $\sim 150000$ galaxies, whose median redshift is $z_{\text{med}}=0.28$.
\par
The background sample has been selected from the sources detected by the Herschel space observatory \citep{PIL10} in the three GAMA fields, covering a total area of $\sim 147\, \text{deg}^2$, and the part of the South Galactic Pole (SGP) that overlaps with the foreground sample ($\sim 60 \,\text{deg}^2$). To ensure no overlap in the redshift distributions of lenses and background sources, we selected only background sources with photometric redshift $1.2 < z < 4.0$. The redshift estimation is described in \cite{gonzaleznuevo17} and \cite{bonavera19}. After performing such a selection, we end up with 57930 galaxies, approximately 24\% of the initial sample.
\begin{figure}[ht]
\centering
\includegraphics[width=\columnwidth]{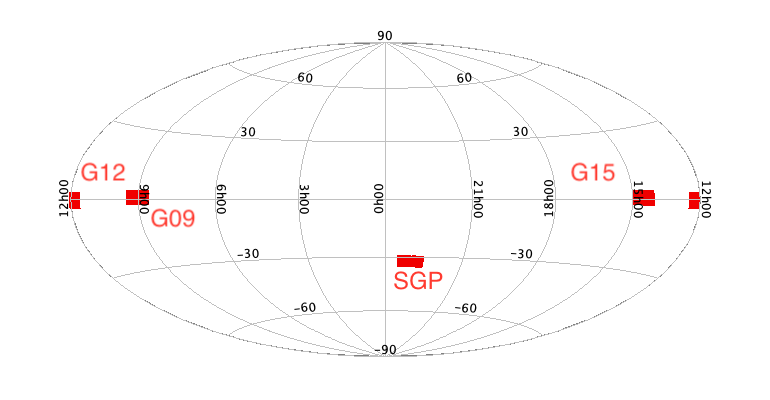}\\
\includegraphics[width=\columnwidth]{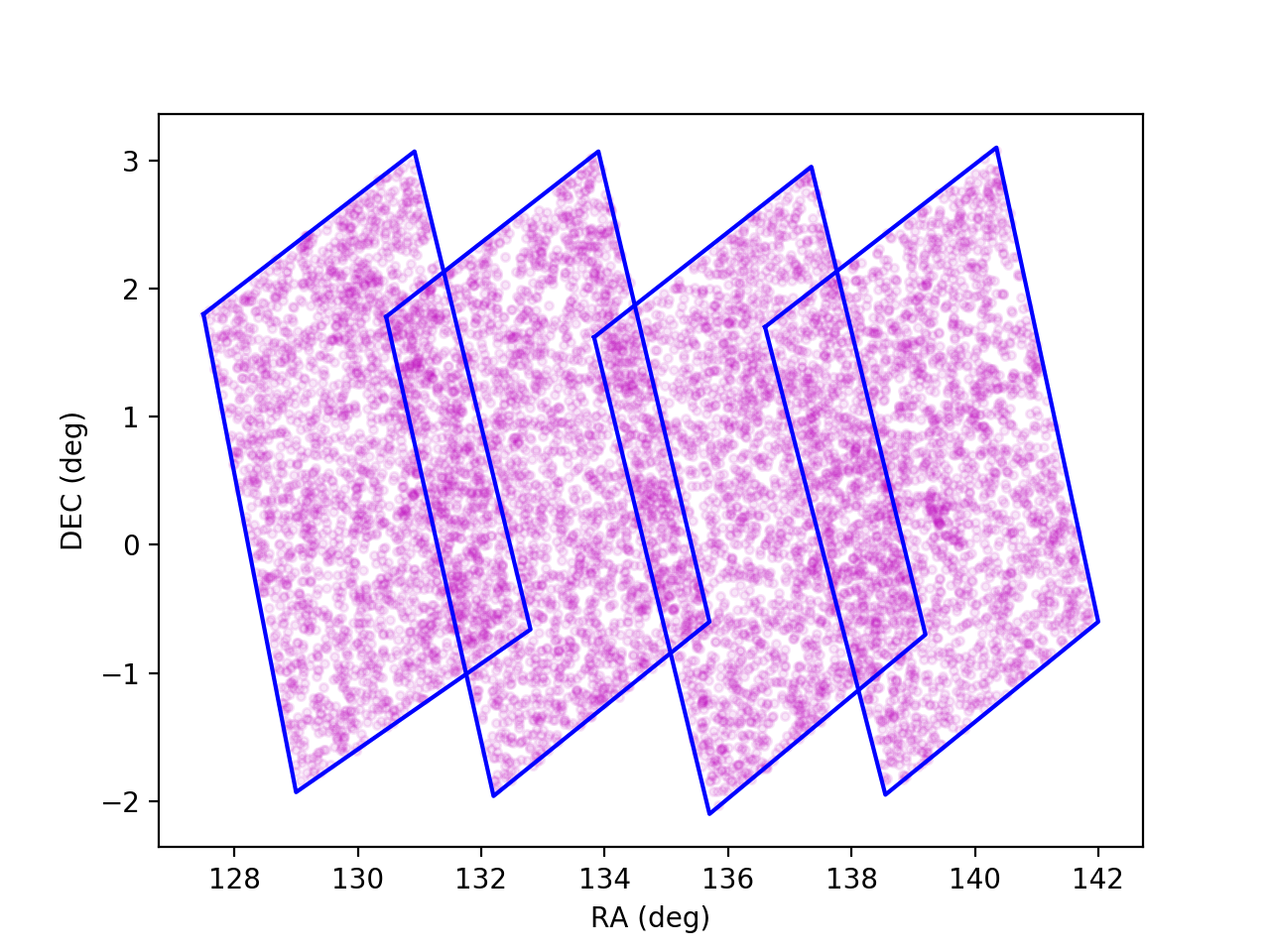}
\caption{Description of the surveyed areas and tiling scheme. Top panel: Mollweide view of the sky distribution of the G09, G12, G15 and SGP regions in equatorial coordinates. Bottom panel: Representation of the Tiles scheme for G09, the pattern being similar for the other regions.}
\label{Fig:tiles}
\end{figure}
\par
It should be stressed that both the H-ATLAS and the GAMA II surveys were carried out to maximize the common area coverage. Both surveys covered the three equatorial regions at 9, 12, and 14.5 h (referred to as G09, G12 and G15, respectively) and the H-ATLAS SGP was also partially observed by GAMA II. Thus, the resulting common area is of about $\sim 207 \text{deg}^2$, surveyed down to a limit of $r \simeq 19.8$ mag. Figure~\ref{Fig:tiles} (top panel) highlights the distributions of the G09, G12, G15, and SGP regions on a Mollweide projection of the sky in equatorial coordinates.

\begin{figure}[ht]
\centering
\includegraphics[width=\columnwidth]{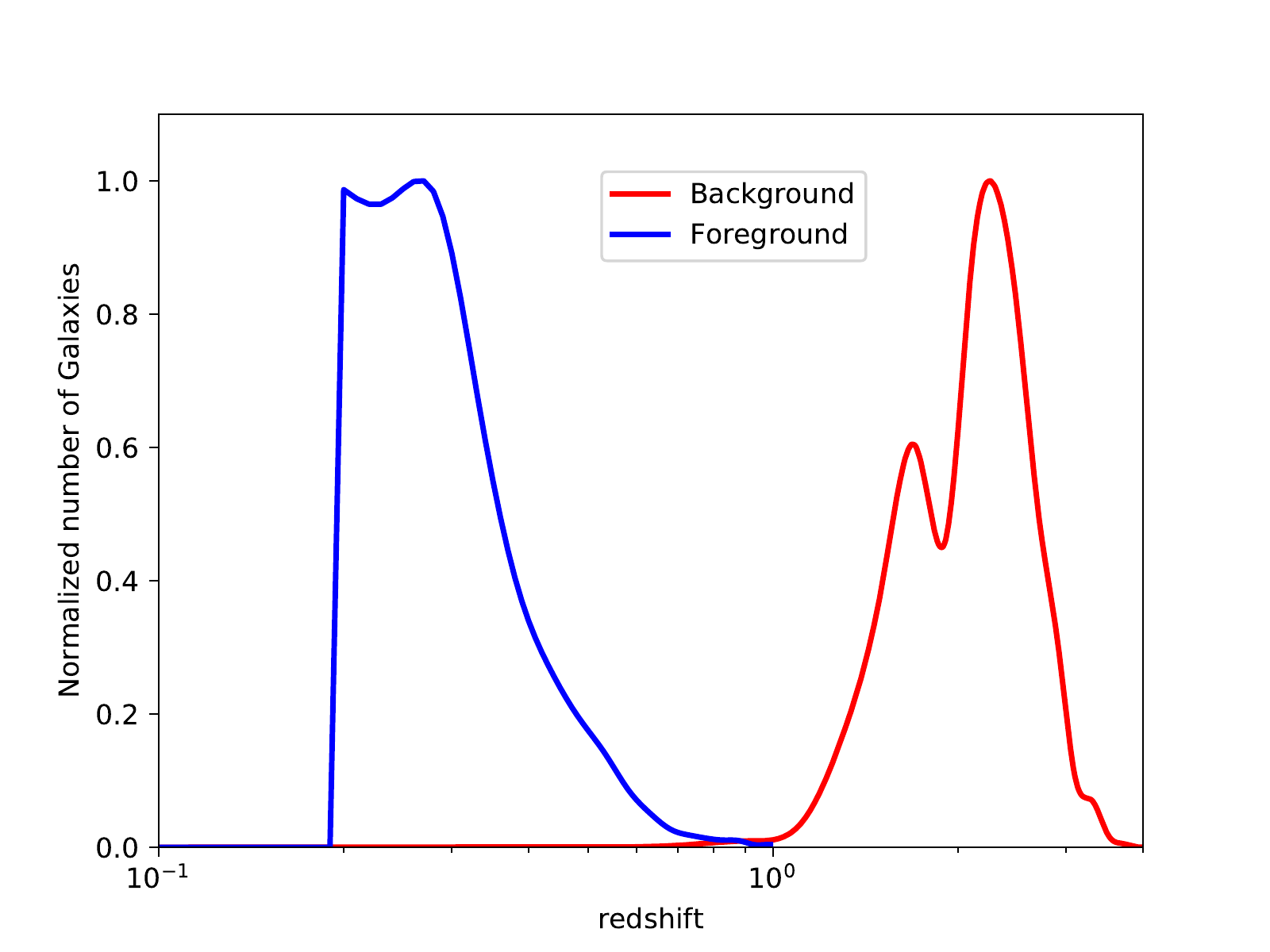}
\caption{Normalized redshift distribution of the background H-ATLAS sample (red) and the foreground GAMA one (blue).}
\label{Fig:dNdz_data}
\end{figure}

Figure \ref{Fig:dNdz_data} shows the normalized redshift distribution of the background and foreground samples (red and blue lines, respectively). This redshift distribution is the estimated $p(z|W)$ of the galaxies selected by our window function and takes into account the effect of random errors in photometric redshifts, as in \cite{gonzaleznuevo17, bonavera20}.

%3.2.Measurements
\subsection{Measurements}
\label{Measurements}
The H-ATLAS survey is divided into five different fields: three GAMA fields in the ecliptic (9h, 12h, 15h) and two in the North and South Galactic Poles (NGP and SGP). The H-ATLAS scanning strategy produced a characteristic repeated diamond shape in most of their fields that was named "Tiles." The area of each tile is $\sim16\,\text{deg}^2$.  In order to maintain a regular shape for the tiles, a small overlap among such regions is needed, typically lower than 20\% of their area. Considering the common area between foreground and background surveys, we have 16 different tiles, which helps diminish the effects of cosmic variance. In particular, Figure~\ref{Fig:tiles} (bottom panel) illustrates the diamond-shaped Tiles scheme in the G09 region. The other considered regions have an analogous pattern.

In this work, we use the angular cross-correlation function measured by \citet{GON20} using the Tiles area for the same spectroscopic sample. We chose this particular set of measurements based on the analysis performed by \citet{GON20}, which studied the large-scale biases for different samples and tiling schemes. The measurements from the spectroscopic sample are only affected by the so called integral constraint \citep[IC;][]{roche99}, but the correction for the chosen tiling scheme is almost negligible (IC $= 5\cdot10^{-4}$). It affects only marginally the measurements at the largest angular scales.

For completeness, we summarize here the pipeline used to estimate the measured cross-correlation function (black circles in Figures \ref{postsampling_ST_p005} and \ref{postsampling_tinker}). As described in detail in \citet{gonzaleznuevo17}, we used a modified version of the \cite{landy93} estimator \citep{herranz01}:
\begin{equation}
\label{eq:wx}
\tilde{w}(\theta)=\frac{\rm{D}_f\rm{D}_b(\theta)-\rm{D}_f\rm{R}_b(\theta)-\rm{D}_b\rm{R}_f(\theta)+\rm{R}_f\rm{R}_b(\theta)}{\rm{R}_f\rm{R}_b(\theta)},
\end{equation}
where $\rm{D}_f\rm{D}_b$, $\rm{D}_f\rm{R}_b$, $\rm{D}_b\rm{R}_f$ and $\rm{R}_f\rm{R}_b$ are the normalized foreground-background, foreground-random, background-random and random-random pair counts for a given separation $\theta$.

The cross-correlation is computed for each tile and its statistical error is obtained by averaging over 10 different realizations (using different random catalogs each time). The final cross-correlation measurement for a given angular separation bin corresponds to the mean value of the cross-correlation functions estimated for every tile. The associated uncertainty is the standard error of the mean, that is, $\sigma_\mu=\sigma/\sqrt{n}$, with $\sigma$ the standard deviation of the population and $n$ the number of independent areas (each selected region can be assumed to be statistically independent due to the small overlap between the tiles).

%3.3. Parameter estimation
\subsection{Parameter estimation}
\label{sec:par_est}

The estimation of the HMF parameters will be carried out through an MCMC method using the open source \emph{emcee} software package \citep{foreman13}, a Python implementation of the Goodman \& Weare affine invariant MCMC ensemble sampler \citep{goodman10}.
\par
As described in Section 2.1, we will adopt two different fits for the HMF. Assuming Gaussian errors, the log-likelihood function takes the form
\begin{align*}
    \log\mathcal{L}(\theta_1,\ldots,\theta_n;\{p_j\}_j)=-\frac{1}{2}\sum_{i=1}^{n}&\bigg[\log{2\pi\sigma_i^2}+\\
    &+\frac{[w(\theta_i;\{p_j\}_j)-\tilde{w}(\theta_i)]^2}{\sigma_i^2}\bigg],
\end{align*}
where $\{p_j\}_j$ is the set of HMF parameters, $\sigma_i$ is the error in the $i_{\text{th}}$ measurement and $w(\theta_i)$ and $\tilde{w}(\theta_i)$ are the theoretical and measured value of the cross-correlation at angular scale $\theta_i$.
\par
With regard to the choice of priors, we consider it a delicate issue. We opted for uniform distributions for all HMF parameters, but the range of these intervals is not obvious at first sight. Furthermore, while some parameters are mathematically forced to be nonnegative ($a_1$ in the ST fit and $A_2$, $B_2$ and $C_2$ in Tinker's), others could a priori be allowed to be negative ($p_1$ in the ST fit and $p_2$ in Tinker's). Traditional methods to determine the HMF imply using an optimizer to find the single tuple of parameter values that best fits the simulations through a $\chi
^2$ analysis and provide no information about whether negative values were allowed in the search. In fact, we have found no mention whatsoever to the potential non-positivity of any of the parameters. As a consequence, for example, while previous simulation-based fits have yielded a value of $p_1\approx 0.3$ for the ST fit, we do not think there is a physically motivated reason to exclude negative values from its priors. As a consequence, even though the main cases we have performed assume all HMF parameters are positive, we also decided to consider the non-negativity of $p_1$, as we will discuss in Section 5 together with the possibility of not applying the normalization condition \eqref{normalization} to the ST fit.
\par
Concerning the HOD parameters, for the runs in which we keep them fixed, we selected the following values based on the \citet{bonavera20} results:
\begin{equation*}
    \alpha=0.9\quad\quad \log{M_{\text{min}}}=12.4\quad\quad \log{M_1}=13.6,
\end{equation*}
where $M_{\text{min}}$ and $M_1$ are expressed in $M_{\odot}/h$, while the Gaussian distributions for the runs in which we include them are extracted from recent literature, as described in \cite{bonavera20}. In particular, they are based on \citet{SIF15} (making use of the recipe by \citet{PAN19} to switch from stellar mass $M_\star$ to halo mass $M_h$) for $M_{min}$ and $M_1$ (in agreement with \citet{AVE15} for $M_{min}$), and on \citet{VIO15} for $\alpha$.
\par
Therefore, the main MCMC runs, along with their respective prior distributions are the following: Run 1 analyzes the two-parameter ST fit with uniform priors, $\mathcal{U}$, on $a_1$ and $p_1$ and fixed HOD parameters, that is,
\begin{equation*} a_1{\sim}\mathcal{U}[0,10]\quad p_1{\sim}\mathcal{U}[0,0.5].
\end{equation*}
Run 2 studies the two-parameter ST fit with uniform priors on $a_1$ and $p_1$ and Gaussian,  $\mathcal{N}$, priors on the HOD parameters:
\begin{align*} a_1{\sim}\mathcal{U}[0,10]\quad p_1{\sim}\mathcal{U}[0,0.5]\quad \alpha{\sim}\mathcal{N}(0.92,0.15)\\
\log{M_{\text{min}}}{\sim}\mathcal{N}(12.4,0.1)\quad \log{M_1}{\sim}\mathcal{N}(13.95,0.3).
\end{align*}
Run 3 analyzes the four-parameter Tinker-like fit with uniform priors on $A_2$, $B_2$, $C_2$ and $p_2$ and fixed HOD parameters:
    \begin{equation*}
       A_2{\sim}\mathcal{U}[0,5]\quad B_2{\sim}\mathcal{U}[0,5]\quad C_2{\sim}\mathcal{U}[0,5] \quad p_2{\sim}\mathcal{U}[0,5].
    \end{equation*}
Run 4 studies the four-parameter Tinker-like fit with uniform priors on $A_2$, $B_2$, $C_2$ and $p_2$ and Gaussian priors on the HOD parameters:
\begin{align*}
       A_2{\sim}\mathcal{U}[0,5]\quad B_2{\sim}\mathcal{U}[0,5]\quad C_2{\sim}\mathcal{U}[0,5] \quad p_2{\sim}\mathcal{U}[0,5] \\ 
       \alpha{\sim}\mathcal{N}(0.92,0.15)\quad
       \log{M_{\text{min}}}{\sim}\mathcal{N}(12.4,0.1)\\ \log{M_1}{\sim}\mathcal{N}(13.95,0.3)
\end{align*}
Lastly, we also performed three additional runs in order to study the possibility of nonpositive values for $p_1$ and not normalizing the ST fit. We will describe them in detail in Section 5.
%----------------
%----------------
%----------------
%4. RESULTS
%----------------
%----------------
%----------------
\section{Main results}

%---------------------------
%--------------------------
%4.1. Sheth and Tormen fit
%--------------------------
%----------------------------
\subsection{Sheth and Tormen function}

%Table run 1
\begin{table*}[t]
\caption{Parameter priors and marginalized posterior peaks, means, $68\%$, and $95\%$ credible intervals for run 1 of the MCMC algorithm, that is, a two-parameter ST fit with positive $p_1$ and fixed HOD values.}
\centering
\begin{tabular}{c c c c c c c}
\hline
\hline
Parameter & Prior & Peak & Mean & $68\%$ CI& $95\%$ CI\\
\hline 
$a_1$&$\mathcal{U}$[0,10]&$0.88$&$1.29$&$[0.42,1.53]$&$[0.10,3.37]$\\
$p_1$&$\mathcal{U}$[0,0.50]&$\quad-\quad$&$0.13$&$[\,\,\,-\,\,\,,0.17]$&$[\,\,\,-\,\,\,,0.31]$\\
\hline
\hline
\end{tabular}
\label{table_ST_p005}
\end{table*}

%Table run 2
\begin{table*}[t]
\caption{Parameter priors and marginalized posterior peaks, means, $68\%$, and $95\%$ credible intervals for run 2 of the MCMC algorithm, that is, a two-parameter ST fit with positive $p_1$ and Gaussian priors on the HOD parameters. Parameters $M_{\text{min}}$ and $M_1$ are expressed in $M_{\odot}/h$.}
\centering
\begin{tabular}{c c c c c c c}
\hline
\hline
Parameter & Prior & Peak & Mean & $68\%$ CI& $95\%$ CI\\
\hline 
$a_1$&$\mathcal{U}$[0,10]&$1.58$&$1.88$&$[0.87,2.42]$&$[0.25,3.72]$\\
$p_1$&$\mathcal{U}$$[0,0.50]$&$0.07$&$0.15$&$[\,\,\,-\,\,\,,0.20]$&$[\,\,\,-\,\,\,,0.33]$\\
$\alpha$&$\mathcal{N}$$[0.92,0.15]$&$0.94$&$0.95$&$[0.80,1.09]$&$[0.67,1.23]$\\
$\text{log}{M}_{\text{min}}$&$\mathcal{N}$$[12.40,0.10]$&$12.48$&$12.46$&$[12.37,12.57]$&$[12.26,12.65]$\\
$\text{log}{M}_{1}$&$\mathcal{N}$[13.95,0.3]&$12.74$&$13.03$&$[12.44,13.26]$&$[12.24,14.27]$\\
\hline
\hline
\end{tabular}
\label{table_ST_gaussian}
\end{table*}

%Posterior sampling runs 1 and 2
\begin{figure*}[ht]
\centering
\includegraphics[width=0.46\textwidth,height=6cm]{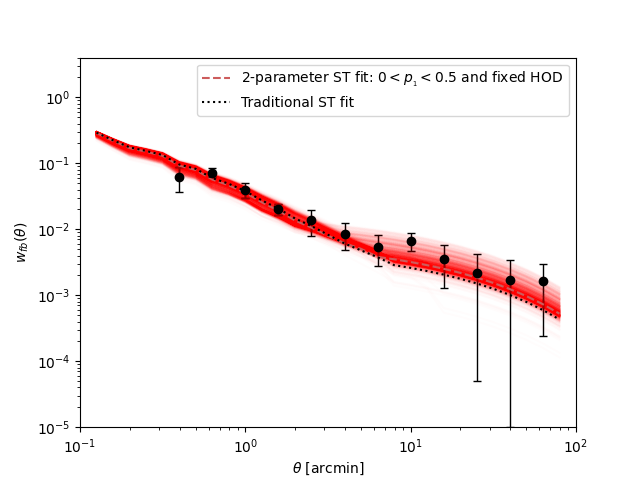}
\includegraphics[width=0.46\textwidth,height=6cm]{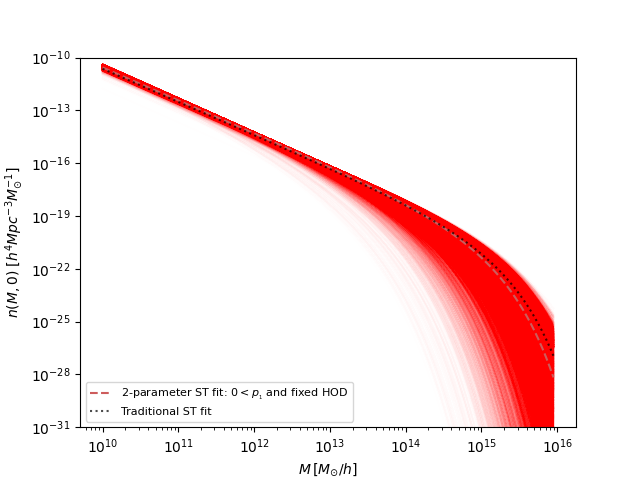}
\includegraphics[width=0.46\textwidth,height=6cm]{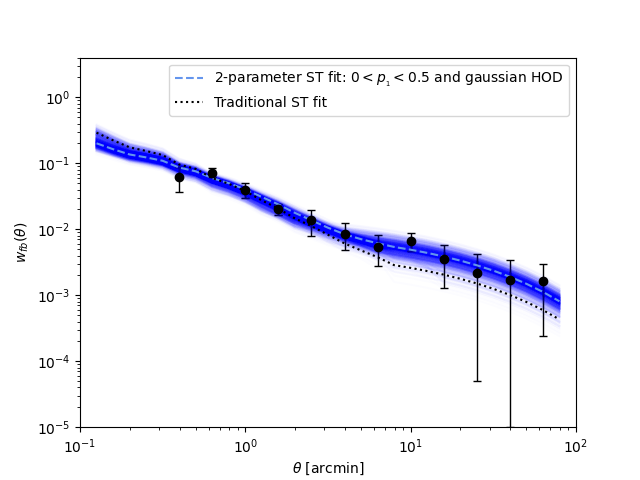}
\includegraphics[width=0.46\textwidth,height=6cm]{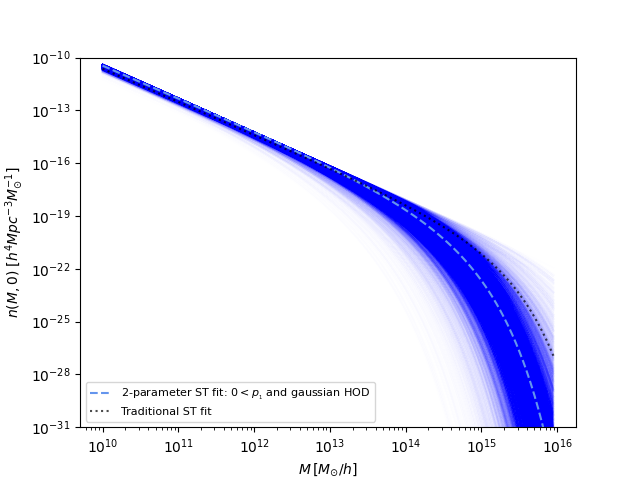}
\caption{Full posterior sampling (solid lines) and mimic marginal mode values (dashed lines) for runs 1 (in red) and 2 (in blue) of the MCMC algorithm, that is, a two-parameter ST fit with fixed and Gaussian priors on the HOD values, respectively. The left panels show the cross-correlation function (the black filled circles being our measurements), while the right panels display the $z=0$ HMF. In all four panels, the dotted black line corresponds to the traditional ST fit.} 
\label{postsampling_ST_p005}
\end{figure*}

Table \ref{table_ST_p005} shows the results from the first run of the MCMC algorithm, namely the peaks, means and narrowest $68\%$ and $95\%$ credible intervals of the marginalized one-dimensional distributions. Figure \ref{cornerplot_ST_fixedvsgaussian} (in blue) shows the corner plot with the one-dimensional and two-dimensional posterior distributions of both parameters. While $a_1$ presents a constraining marginalized posterior with a clear peak at $a_1=0.88$, $p_1$ can only be assigned upper bounds, namely $p_1<0.17$ and $p_1<0.31$ at 68\% and 95\% credibility, respectively. For our fixed choice of HOD parameters, the traditional parameter values of the ST fit are compatible given the wide uncertainties in the posterior distributions, although the marginal mean value of $p_1$ hints at smaller values.

The upper-left panel of Figure \ref{postsampling_ST_p005} shows the resulting cross-correlation function when the full posterior distribution is sampled (solid red lines), along with the lines corresponding to the traditional ST fit (dotted black) and the "mimic" marginal peak values (dashed light red), corresponding to $a_1=0.88$ and $p_1=0.20$. Since the marginalized posterior of $p_1$ does not display a peak, the latter line has been chosen so that it provides a reasonable fit and serves only as a visual aid, hence the word mimic. As can be seen from its comparison to the measured data (black circles), there is more probability density toward smaller cross-correlation values at angular scales $\theta>3$ arcmin: The model does not appear to be able to fully explain the large-scale data.
\par
The $z=0$ HMF corresponding to the sampling of the full posterior distribution (solid red lines) is plotted in the upper-right panel of Figure \ref{postsampling_ST_p005} and compared with the traditional ST fit (dotted black line). Our results are compatible within the uncertainties, although there appears to be a tendency toward a smaller number of halos at large masses ($M>10^{13.5} M_{\odot}/h$). At low masses, the HMF is well-constrained, whereas our treatment provides interesting upper bounds for the HMF at the aforementioned large scale.

As expected, if we now introduce the HOD parameters in the MCMC analysis (with Gaussian priors as discussed in Section \ref{sec:par_est}), the results, which we present in Table \ref{table_ST_gaussian} and Figure \ref{cornerplot_ST_fixedvsgaussian} (in red) vary quantitatively. With respect to the fixed HOD case, the $a_1$ and $p_1$ marginalized distributions present some differences. In particular, both the peak and the mean of the $a_1$  distribution are displaced to the right to values of 1.58 and 1.88, respectively. Moreover, the $p_1$ distribution, while still right-skewed, becomes mainly concave with a mode of $p_1=0.07$, as opposed to the first run. Concerning the HOD parameters, whereas the marginalized posterior distributions of $\alpha$ and $\log{M_{\text{min}}}$ hardly deviate from their priors (with peaks at 0.94 and 12.48, respectively), that of $\log{M_1}$ does substantially, with a clear peak at 12.74, more than $3\sigma$ away from its prior mean. 

The lower-left and lower-right panels of Figure \ref{postsampling_ST_p005} show the corresponding posterior-sampled cross-correlation and $z=0$ HMF (solid blue lines) along with the traditional ST fit (dotted black line) and the marginal peak values (dashed light blue). The introduction of the HOD parameters in the MCMC analysis has now allowed the model to properly explain the large-scale data. Compared to the previous case, the derived HMF hints at a general tendency toward fewer halos, notably at masses $M>10
^{13.4} M_{\odot}/h$.

In summary, when the HOD parameters are fixed, the two-parameter ST fit is not able to fully explain the cross-correlation signal at angular scales $\theta>3$ arcmin. Although a larger value of $a_1$ would help in this direction (as a parameter sensitivity analysis shows), this would provide a poorer general fit to the data because it would cause the small-scale cross-correlation, which is better constrained by observations, to decrease. It should be noted that the role of $p_1$ is not as significant in this argument given the little room for manoeuvre (prior-wise) at its disposal. 

However, the situation differs for the case in which the HOD parameters are introduced in the MCMC analysis. As described in \cite{bonavera20}, a decrease in parameter $M_1$ mainly causes an increase in the cross-correlation function, this effect being more noticeable at angular scales between 1 and 4 \mbox{arcmin} and almost negligible at larger scales. As a consequence, $a_1$ can now be increased in order to accommodate the data without impoverishing the fit by demanding that $M_1$ be decreased, that is, that there be more satellite galaxies. The sampling of the full posterior (lower-left panel of Figure \ref{postsampling_ST_p005}) reflects this situation clearly. It should also be mentioned that larger values of $M_{\text{min}}$ have an increasing effect on all scales, again to the detriment of smaller-scale values and thus diminishing its influence.

Although the posterior distribution for $M_1$ is physically reasonable, it differs substantially from those obtained by \cite{bonavera20} or \cite{GON20} using the traditional ST fit, which should serve as additional motivation for the analysis in Section 5. In any event, as compared to the traditional one, the ST fit as described in this section hints at a smaller number of halos, especially for the largest masses, an effect that is mainly driven by the cross-correlation measurements at $\theta>3$ arcmin.

%----------------
%Tinker fit
%---------------
\subsection{Tinker-like function}

%Table run 3
\begin{table*}[ht]
\caption{Parameter priors and marginalized  posterior peaks, means, $68\%$, and $95\%$ credible intervals for run 3 of the MCMC algorithm, that is, a four-parameter Tinker fit and fixed HOD values.}
\centering
\begin{tabular}{c c c c c c}
\hline
\hline
Parameter & Prior & Peak & Mean & $68\%$ CI& $95\%$ CI\\
\hline 
$A_2$&$\mathcal{U}$[0,5]&0.15&$0.20$&$[0.08,0.29]$&$[0.02,0.38]$\\
$B_2$&$\mathcal{U}$[0,5]&$0.82$&$1.66$&$[\,\,\,-\,\,\,,1.96]$&$[\,\,\,-\,\,\,,\,\,\,-\,\,\,]$\\
$C_2$&$\mathcal{U}$[0,5]&0.56&$0.78$&$[0.33,1.00]$&$[0.15,1.55]$\\
$p_2$&$\mathcal{U}$[0,5]&$-$&$-$&$[\,\,\,-\,\,\,,\,\,\,-\,\,\,]$&$[\,\,\,-\,\,\,,\,\,\,-\,\,\,]$\\
\hline
\hline
\end{tabular}
\label{table_tinker}
\end{table*}

%Table run 4
\begin{table*}[ht]
\caption{Parameter priors and marginalized posterior peaks, means, $68\%$, and $95\%$ credible intervals for run 4 of the MCMC algorithm, that is, a four-parameter Tinker fit with Gaussian priors on the HOD parameters. Parameters $M_{\text{min}}$ and $M_1$ are expressed in $M_{\odot}/h$.}
\centering
\begin{tabular}{c c c c c c c}
\hline
\hline
Parameter & Prior & Peak & Mean & $68\%$ CI& $95\%$ CI\\
\hline 
$A_2$&$\mathcal{U}$[0,5]&$0.16$&$0.20$&$[0.07,0.28]$&$[\,\,\,-\,\,\,,0.39]$\\
$B_2$&$\mathcal{U}$[0,5]&$0.91$&$1.71$&$[\,\,\,-\,\,\,,2.04]$&$[\,\,\,-\,\,\,,\,\,\,-\,\,\,]$\\
$C_2$&$\mathcal{U}$[0,5]&$0.63$&$0.85$&[0.27,1.11]&[0.01,1.80]\\
$p_2$&$\mathcal{U}$[0,5]&$-$&$-$&$[\,\,\,-\,\,\,,\,\,\,-\,\,\,]$&$[\,\,\,-\,\,\,,\,\,\,-\,\,\,]$\\
$\alpha$&$\mathcal{N}$[0.92,0.15]&$0.89$&$0.91$&[0.77,1.06]&[0.62,1.21]\\
$\text{log}{M}_{\text{min}}$&$\mathcal{N}$[12.40,0.10]&$12.43$&$12.42$&[12.32,12.42]&[12.23,12.62]\\
$\text{log}{M}_{1}$&$\mathcal{N}$[13.95,0.30]&$12.91$&$13.56$&[12.58,14.20]&[12.48,14.97]\\
\hline
\hline
\end{tabular}
\label{table_tinker_gaussian}
\end{table*} 

%Posterior sampling run 3
\begin{figure*}[ht]
\centering
\includegraphics[width=0.46\textwidth,height=6cm]{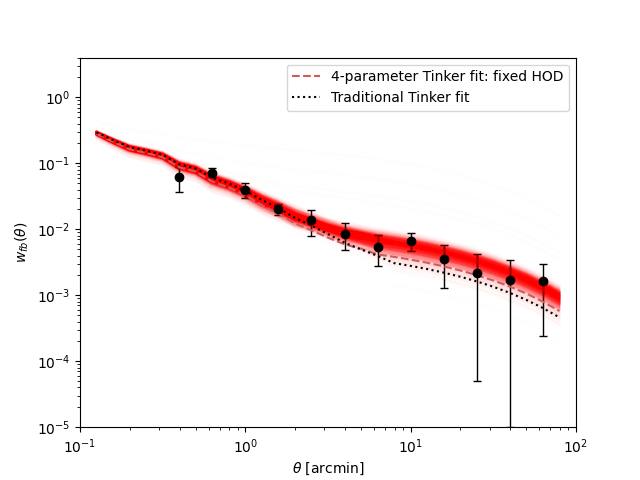}
\includegraphics[width=0.46\textwidth,height=6cm]{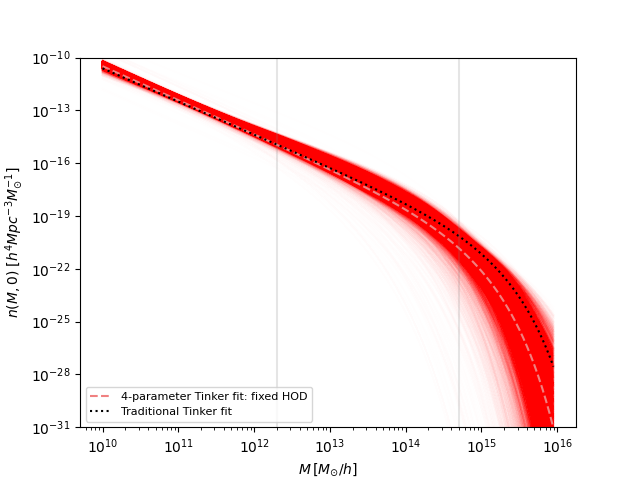}
\includegraphics[width=0.46\textwidth,height=6cm]{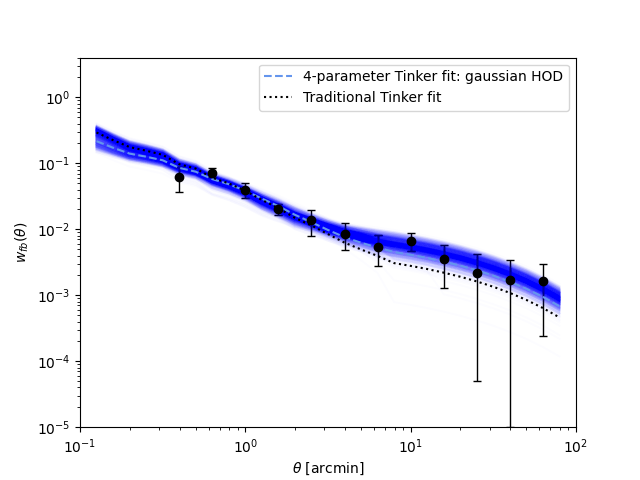}
\includegraphics[width=0.46\textwidth,height=6cm]{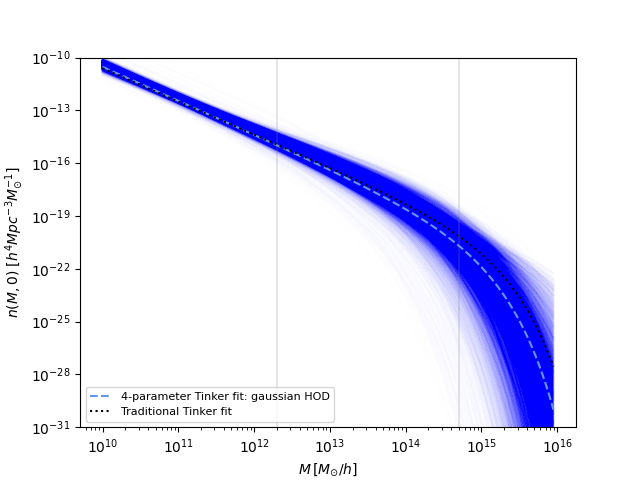}

\caption{Full posterior sampling (solid lines) and mimic marginal mode values (dashed lines) for runs 3 (in red) and 4 (in blue) of the MCMC algorithm, that is, a four-parameter Tinker fit with fixed and Gaussian priors on the HOD values, respectively. The left panels show the cross-correlation function (the black filled circles being our measurements), while the right panels display the $z=0$ HMF. In all four panels, the dotted black line corresponds to the traditional Tinker $z=0$ fit.} 
\label{postsampling_tinker}
\end{figure*}

Table \ref{table_tinker} and Figure \ref{cornerplot_tinker_fixedvsgaussian} (in blue) show the corresponding results for the third run of the MCMC algorithm: a four-parameter Tinker fit with fixed HOD values. Whereas $A_2$ and $C_2$ show constraining marginalized posterior distributions with peaks at $A_2=0.15$ and $C_2=0.56$, $B_2$ and $p_2$ remain unconstrained, the former hinting toward a peak value of 0.91 and the latter being completely prior-dominated. This issue is not resolved by widening the priors (even considering negative values for $p_2$) and therefore compromises the reliability of the statistical conclusions, since the credible intervals on the HMF will eventually depend on the prior range of $B_2$ and $p_2$. However, we suspect that the derived HMF is not too sensitive to this issue, although we consider it delicate and have not gone further into a quantitative analysis. It should also be added that parameters $A_2$ and $C_2$ are very robust to the widening or narrowing of said prior distributions.

The red lines in Figure \ref{postsampling_tinker} show the posterior-sampled cross-correlation function (upper-left panel) and $z=0$ HMF (upper-right panel), along with the traditional Tinker fit 
(dotted black line) and the mimic marginal mode values (dashed faint red line), corresponding to $A_2=0.15$, $B_2=0.82$, $C_2=0.56$ and $p_2=1.50$. As opposed to the ST fit, the vast majority of the sampled cross-correlation lines seem to properly explain the large-scale data, while the traditional Tinker fit underestimates the measurements above 1 arcmin. As can be seen from the upper-right panel of Figure \ref{postsampling_tinker}, the $z=0$ HMF is in good agreement with the traditional Tinker fit at the lowest masses and, in particular, for the same mass range used in the derivation of the original Tinker fit. Similarly to the ST case, the derived HMF tends to prefer a steeper cutoff at high masses, although less pronounced. However, the recovered HMF shows a wider spread for low and intermediate masses, $M<10^{14.0}M_\odot/h$, as compared to the previous subsection.

We now turn to analyzing the introduction of the HOD parameters in the MCMC analysis. Table \ref{table_tinker_gaussian} and Figure \ref{cornerplot_tinker_fixedvsgaussian} (in red) show the corresponding results. In this case, the marginalized distributions of the HMF parameters practically show no difference when compared to the fixed HOD case (Table \ref{table_tinker} and Figure \ref{cornerplot_tinker_fixedvsgaussian} in blue). Only the peak and the mean of the $C_2$ distribution are visibly displaced to the right to values of $0.63$ and $0.85$, respectively. Regarding the HOD parameters, the situation resembles that of the previous case up to a certain point; the marginalized posterior distributions of $\alpha$ and $\log{M_{\text{min}}}$ hardly move away from their priors (with peaks at 0.89 and 12.43, respectively), while that of $\log{M_1}$ does (to the left), but in this case appears to maintain a high probability region toward values around the prior.

The lower panels of Figure \ref{postsampling_tinker} show the posterior-sampled (solid blue lines) cross-correlation function (lower-left panel) and $z=0$ HMF (lower-right panel) together with the traditional Tinker fit (dotted black line) and the "mimic" marginal peak values (dashed faint blue line), corresponding to $A_2=0.16$, $B_2=0.91$, $C_2=0.63$ and $p_2=1.50$. Since the data was already properly explained by the fixed HOD case, we only observe an expected increase in the spread of the HMF, mainly in the form of higher upper bounds at every mass, and especially for $M>10
^{14.0} M_{\odot}/h$.

In summary, while the Tinker-like fit shows a more robust behavior with respect to the HOD parameters than the ST fit (due to the fact that, unlike the latter, it can properly explain the cross-correlation signal without changes in them), the statistical results concerning the HMF depend on the prior range of two of its parameters $(B_2$ and $p_2$), which cannot be bounded. Although we do not suspect this is a major issue, it should nevertheless be clarified that the credible intervals derived for the HMF in runs 3 and 4 have assumed the prior ranges described in section 3.3.

%----------------------------------

\section{Further discussion}

%Non-normalization of the HMF
\subsection{Non-normalization of the HMF}

%Posterior sampling 5.1
\begin{figure*}[t]
\centering
\includegraphics[width=0.46\textwidth,height=6cm]{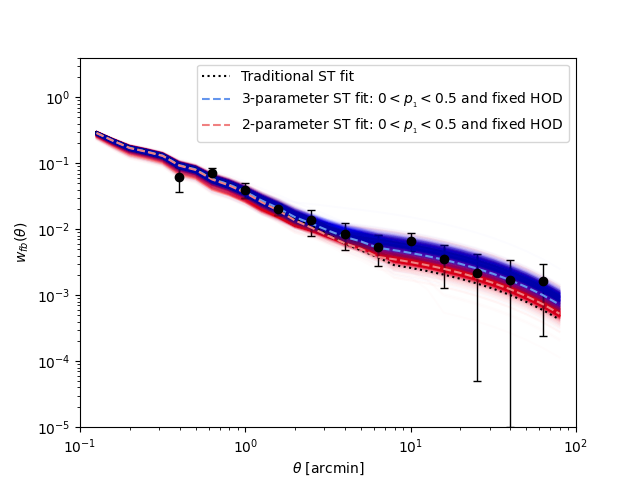}
\includegraphics[width=0.46\textwidth,height=6cm]{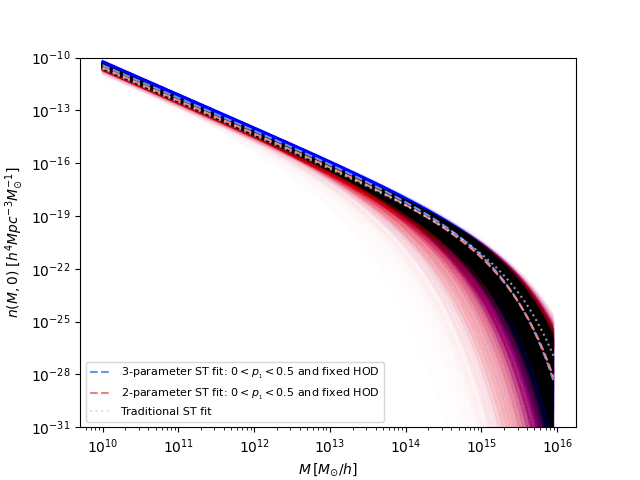}

\caption{Full posterior sampling (solid lines) and (mimic) marginal mode values (dashed lines) from runs 1 (in red) and 5 (in blue) of the MCMC algorithm, that is, a two-parameter fixed HOD and a three-parameter fixed HOD ST fit, respectively. Parameter $p_1$ is assumed to be in the range $[0,0.5)$. The left panel shows the cross-correlation function (the black filled circles being our measurements), while the right one displays the $z=0$ HMF. The dotted black line corresponds to the traditional ST fit. }
\label{comparison_51}
\end{figure*}

%Posterior sampling 5.2
\begin{figure*}[t]
\centering
\includegraphics[width=0.46\textwidth,height=6cm]{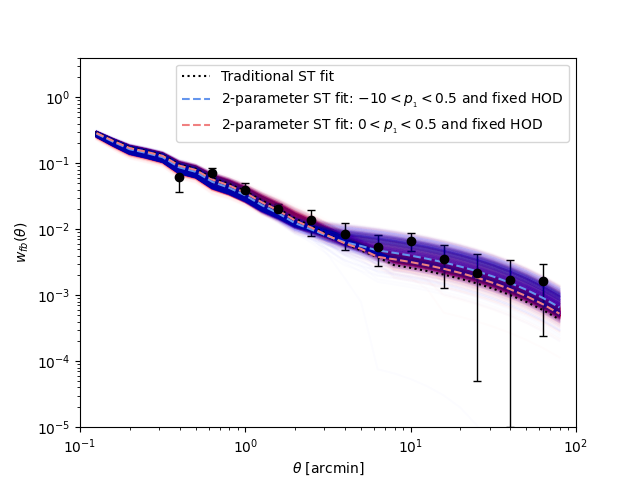}
\includegraphics[width=0.46\textwidth,height=6cm]{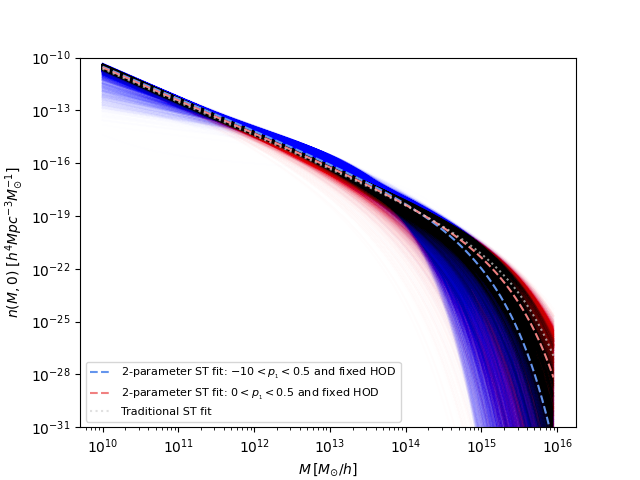}

\caption{Full posterior sampling (solid lines) and (mimic) marginal mode values (dashed lines) from runs 1 (in red) and 6 (in blue) of the MCMC algorithm, that is, a two-parameter fixed HOD ST fit with $p_1>0$ and $p_1$ allowed to be negative, respectively. The left panel shows the cross-correlation function (the black filled circles being our measurements), while the right one displays the $z=0$ HMF. The dotted black line corresponds to the traditional ST fit. }
\label{comparison_52_2parameter}
\end{figure*}

As discussed in Section 2.1, the normalization condition imposed on the ST fit assumes all mass in the Universe is bound up in halos. Although the present work has incorporated this assumption on the grounds of coherence with the underlying halo model, it is of interest to analyze the situation when the $A_1$ parameter is left free in the MCMC analysis. In this scenario, $p_1$ could, in principle, take values that are larger than 0.5 (or even negative; see next subsection) but, for the sake of comparison, we will keep the priors on $a_1$ and $p_1$ the same as in Section 4.1. 

The results are displayed in Table \ref{table_nonnormalization} and Figure \ref{cornerplot_ST_nonnormalization} (in red). Parameter $A_1$ shows a well-constrained marginalized distribution with a peak at $A_1=0.59$, while that of $a_1$ is narrower than that of run 1 and barely displaced to the right, with a peak at $a_1=0.93$. It should be noted that parameter $p_1$ is now unconstrained on both sides, hinting again at a preference for negative values. Figure \ref{comparison_51} further compares the posterior sampling of run 5 with that of run 1. It permits us to conclude that, while keeping $p_1$ positive and smaller than 0.5, the introduction of $A_1$ as a free parameter in the ST fit allows the cross-correlation function to take larger values for $\theta>3$ arcmin, as required by the data, without needing the HOD parameters to vary. This, in turn, translates to a more constrained HMF at large mass values, as can bee seen by the black band in the right panel of Figure \ref{comparison_51}, the area of overlap of both samplings. It should, however, be emphasized that we have assumed the prior range for $p_1$ to be $[0,0.5]$ in order to study the possible qualitative differences with respect to run 1 and to serve as a link between section 4 and the next subsection.

%Non-positivity of HMF parameters
\subsection{Non-positivity of HMF parameters}

%Table run 7
\begin{table*}[t]
\caption{Parameter priors, marginalized  posterior peaks, means, $68\%$, and $95\%$ credible intervals for run 7 of the MCMC algorithm, that is, a three-parameter ST fit, $p_1$ allowed to be negative and fixed HOD values.}
\centering
\begin{tabular}{c c c c c c}
\hline
\hline
Parameter & Prior & Peak & Mean & $68\%$ CI& $95\%$ CI\\
\hline 
$A_1$&$\mathcal{U}$[0,5]&$0.66$&$0.55$&$[0.36,0.87]$&$[\,\,\,-\,\,\,,\,\,\,-\,\,\,]$\\
$a_1$&$\mathcal{U}$[0,10]&$1.29$&$1.30$&$[0.74,2.55]$&$[0.36,3.86]$\\
$p_1$&$\mathcal{U}$[-10,10]&$-1.25$&$-1.15$&$[-2.36,0.11]$&$[-3.50,1.15]$\\
\hline
\hline
\end{tabular}
\label{table_pneg_nonnormalization}
\end{table*}

%Posterior sampling 5.3
\begin{figure*}[ht]
\centering
\includegraphics[width=0.45\textwidth,height=5.8cm]{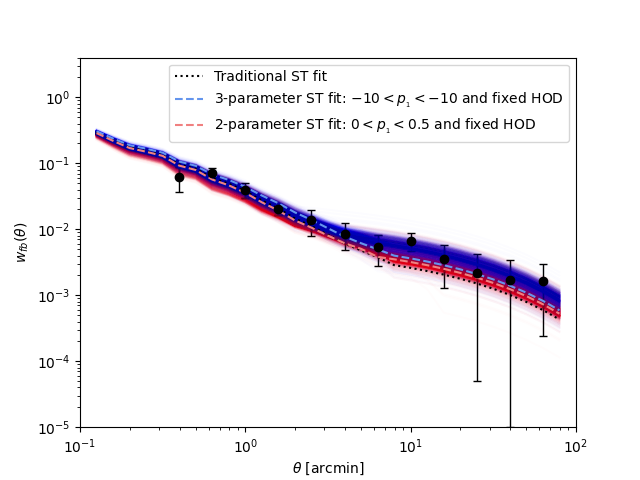}
\includegraphics[width=0.45\textwidth,height=5.8cm]{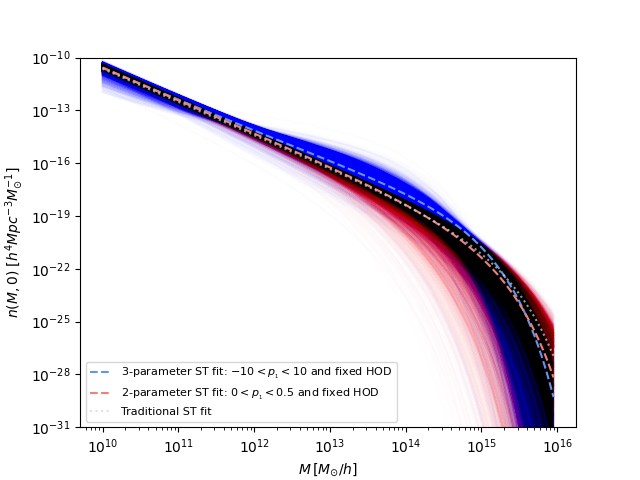}

\caption{Full posterior sampling (solid lines) and (mimic) marginal mode values (dashed lines) from runs 1 (in red) and 7 (in blue) of the MCMC algorithm, that is, a two-parameter fixed HOD ST fit with $p_1>0$ and a three-parameter fixed HOD ST fit with $p_1$ allowed to be negative, respectively. The left panel shows the cross-correlation function (the black filled circles being our measurements), while the right one displays the $z=0$ HMF. The dotted grey line corresponds to the traditional ST fit.}
\label{comparison_52_3parameter}
\end{figure*}

Another point regarding the HMF parameters was raised in Section 3.3. We find no mathematical reason why parameters $p_1$ and $p_2$ cannot take negative values. As to a possible physical explanation, an analysis of the excursion set formalism or of other works that derive a HMF template purely from physical arguments still yields no reason why this cannot be the case. We would like to emphasize that the usual methods consist in finding the single tuple of parameters that provides the best fit to the simulation in question, but we have found no further details about the range of parameter values that is used in said searches (are negative values explored?). Since prior distributions are of paramount importance in Bayesian statistics, we deem this a delicate issue. As a consequence, we decided to analyze the possibility of allowing parameter $p_1$ in the ST fit to take negative values, both in the case where the normalization condition is applied (two-parameter fit) and in the case where it is not (three-parameter fit). The results for both cases are shown in Figure \ref{cornerplot_ST_pneg} and the statistical results are summarized in Tables \ref{table_pneg} and \ref{table_pneg_nonnormalization}, respectively.

In the two-parameter case (Figure \ref{cornerplot_ST_pneg} in blue), we now observe clear peaks in both parameters, at values $a_1=1.46$ and $p_1=-0.43$, and a strong degeneracy direction that produces the appearance of long tails in both one-dimensional marginalized distributions. Figure \ref{comparison_52_2parameter} shows the posterior-sampled cross correlation function (left panel) and $z=0$ HMF (right panel) of run 6 (in blue) compared to that of run 1 (in red). 
From the left panel, we can infer that allowing negative values of $p_1$ helps to account for the high correlation at large angular scales ($\theta>3$ arcmin). However, it is not as sufficient as varying the HOD parameters or the normalization parameter $A_1$ in the MCMC analysis, as shows the fainter blue line density in the cross-correlation sampling.
Moreover, the large degeneracy between $a_1$ and $p_1$ translates to a much wider spread in the HMF, which is clearly visible at the smallest and largest mass values. 

On the other hand, the three-parameter case (Table \ref{table_pneg_nonnormalization} and Figure \ref{cornerplot_ST_pneg} in red) presents a very symmetric marginalized posterior distribution for $p_1$ with a clear peak at $p_1=-1.25$. This parameter shows again a degeneracy with $a_1$, although this does not originate one-sided tails in this case. Parameter $A_1$ peaks at $A_1=0.657$, while $a_1$ does at $1.290$. Figure \ref{comparison_52_3parameter} shows the posterior-sampled cross-correlation function (left panel) and $z=0$ HMF (right panel) of run 7 (in blue) compared again with that of run 1 (in red). As opposed to the previous two-parameter case, allowing negative $p_1$ values can clearly explain the cross-correlation data and, as it can be seen in the right panel of Figure \ref{comparison_52_3parameter}, the derived HMF appears to hint at a larger number of halos when compared to run 1, notably in the range $10
^{12}<M<10^{15} M_{\odot}/h$.
\par
Comparing the two results of the three-parameter case, we observe that the peaks for $A_1$ are almost the same in both scenarios (there is only a slight increase in the credible intervals for the nonpositivity case). However, the $a_1$ peak value increases from 0.93 to 1.29, as do the mean and the upper credible interval. This difference clearly arises from the fact that $p_1$ appears to be driven by the data to take negative values and, in turn, $a_1$ has to increase in order to counteract this effect. Unlike the first run, $p_1$ now has a wide enough range within which it can move, hence the constraining posterior distributions. In summary, introducing $A_1$ as a free parameter along with allowing $p_1$ to take negative values allows us to bypass the two problematic aspects that we have encountered in this paper: the long one-sided tails in the $a_1$ and $p_1$ marginalized posterior distributions and the lack of generality in the choice of prior range. 

%Tabulation of the halo mass function
\subsection{Tabulation of the halo mass function}

\begin{figure}[h]
\centering
\includegraphics[width=\columnwidth]{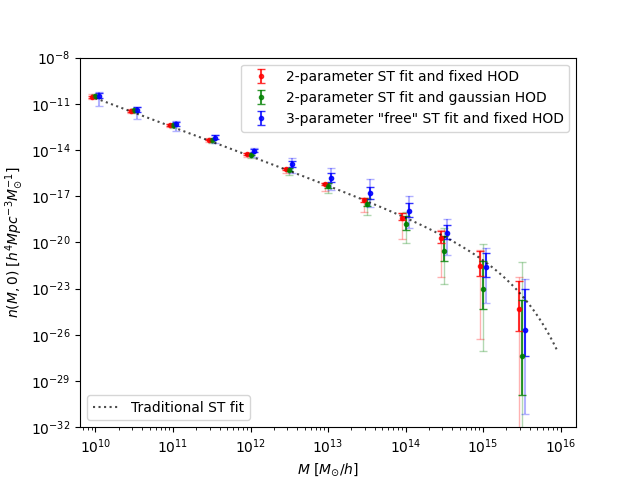}
\caption{Credible intervals (68\% in bold and 95\% in faint colors) for the $z=0$ HMF at different mass values when the full posterior distribution is sampled for the ST fit in the two-parameter fixed HOD case (red), the two-parameter Gaussian HOD case (green), and the three-parameter case (blue). The plots for each case are slightly displaced in the horizontal direction just for visual purposes.}
\label{ST_tabulated_hmf}
\end{figure}

\begin{figure}[h]
\centering
\includegraphics[width=\columnwidth]{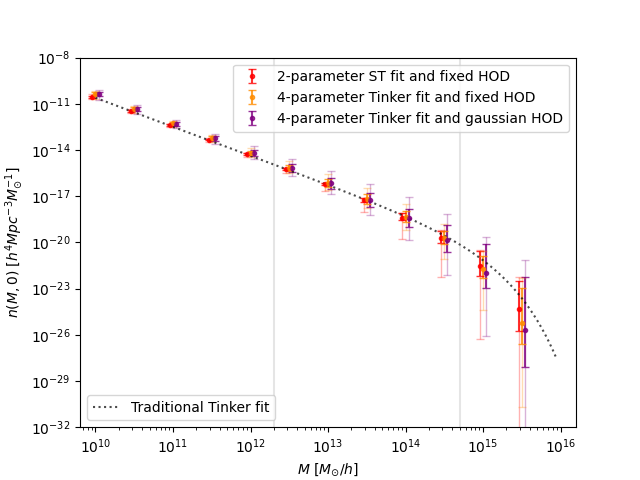}
\caption{Credible intervals (68\% in bold and 95\% in faint colors) for the $z=0$ HMF at different mass values when the full posterior distribution is sampled for the four-parameter Tinker fit in the fixed HOD case (dark orange) and in the Gaussian HOD case (purple). The two-parameter ST fit with fixed HOD is also shown for comparison (red). The plots for each case are slightly displaced in the horizontal direction just for visual purposes.}
\label{tinker_tabulated_hmf}
\end{figure}

With a view to constraining the HMF itself at any redshift (irrespective of its parameters), we now make use of one of the main advantages of performing a Bayesian analysis and study the spread of the full posterior distribution so as to obtain credible intervals for the value of the $z=0$ HMF at given masses. In other words, the information contained in the red and blue bands shown in Figures \ref{postsampling_ST_p005}, \ref{postsampling_tinker} and \ref{comparison_52_3parameter} has been summarized at certain mass values. The resulting plots are shown in Figures \ref{ST_tabulated_hmf} and \ref{tinker_tabulated_hmf}, where the HMF is plotted at mass values ranging from $10
^{10}$ to $10^{15.5}$ $M_{\odot}/h$ for each case. The associated numerical values are tabulated in Tables \ref{tabulationST} and \ref{tabulationtinker}.

Figure \ref{ST_tabulated_hmf} shows the median, 68\%, and 95\% credible intervals for the $z=0$ HMF at different mass values for the two-parameter ST fit with $p_1>0$ and fixed HOD parameters (that is, run 1, in red), the two-parameter ST fit with $p_1>0$ and Gaussian priors on the HOD parameters (run 2, in green) and the three-parameter ST fit with fixed HOD values, meaning the case where $p_1$ is allowed to be negative and greater than 0.5 (run 7, in blue). There is good agreement with the traditional ST fit (black dotted line), with a tendency toward fewer massive halos at mass values larger than $M\gtrsim 10^{14}\,M_{\odot}/h$ in the first two cases. The three-parameter ST fit shows the previously mentioned tendency toward a larger number of halos at intermediate masses, between $10^{11.5}$ and $10^{15}\,M_{\odot}/h$, although still compatible with the traditional ST fit within the uncertainties.

%ST hmf tabulation

Figure \ref{tinker_tabulated_hmf} shows the corresponding results for the four-parameter Tinker fit with $p_2>0$ and fixed HOD parameters (that is, run 3, in orange) and the four-parameter Tinker fit with $p_2>0$ and Gaussian priors on the HOD parameters (run 4, in purple). The two-parameter ST fit with $p_1>0$ and fixed HOD is also depicted in red for comparison. In both cases, there is very good agreement with the traditional Tinker fit (black dotted line) although, as commented in Section 4.2, there is a wider spread for low and intermediate mass values ($M<10
^{14.0} M_{\odot}/h$) when compared to the ST fits.

%Tinker hmf tabulation

%----------------
%----------------
%6. Summary and conclusions
%----------------
%----------------
\section{Summary, conclusions and future prospects}
This paper has explored the submillimeter galaxy magnification bias as a cosmological observable to provide a proof-of-concept method to extract information about the HMF. By means of a halo model interpretation of the foreground-background cross-correlation function between samples of GAMA II (with spectroscopic redshift between $0.2<z<0.8$ and $z_{\text{med}}=0.28$) and H-ATLAS galaxies (with photometric redshift between $1.2<z<4.0$ and $z_{\text{med}}=2.2$), we carried out a Bayesian analysis with two different universal HMF models with the aim of studying which of them provides a better fit to the data and deriving observation-based credible intervals for the number density of the dark matter halos associated with the lenses at certain mass values. We have also studied the potential influence of the HOD parameters in our conclusions. 
\par
We have begun our analysis with the apparently common assumption that all HMF parameters should be positive. In this scenario, we have found that the two-parameter ST fit can only properly explain the cross-correlation signal at angular scales larger than 3 arcmin when the HOD parameters are introduced in the MCMC analysis and thus allowed to vary. Indeed, the two-parameter ST fit is shown to be sensitive to the variation of the HOD and a decrease in $M_1$ (which substantially deviates from its prior distributions) with a corresponding increase in $a_1$ allows it to properly reproduce the data. On the other hand, the four-parameter Tinker fit is quite robust to changes in the HOD parameters and easily accommodates the large-scale data, but two of its parameters cannot be constrained. In fact, the extent of their posterior distributions depends on the corresponding range of their priors and this is a delicate issue when trying to derive statistical results. In other words, care should be taken when interpreting our statistics of the Tinker fit, since they rely on our specific assumption of prior ranges, although we do not suspect major differences would appear if they were modified.

Both cases have nonetheless yielded credible intervals for the $z=0$ HMF that display similar features, in that they are in general agreement with the traditional fits obtained from N-body simulations and constrain the HMF with same-order uncertainties. The Tinker fit, however, appears to hint at a larger number of halos for low and intermediate masses ($M<10^{14.0} M_{\odot}/h)$.

We next analyzed the possibility of relaxing the normalization assumption for the ST fit while keeping parameter $p_1$ within the range $[0,0.5)$ for the sake of comparison with the two-parameter case. We found that, under these conditions, adding $A_1$ as a free parameter in the analysis allows the ST model to properly explain the cross-correlation data at the largest scales without resorting to changes in the HOD parameters. Parameter $p_1$, however, now becomes unconstrained on both ends, which serves as a hint that other values should be explored.

Indeed, motivated by the large relevance of prior distributions in Bayesian inference and by the impossibility of constraining parameter $p_1$ on both sides with the previous studies, we decided to consider the case of a wide enough prior range for it, since we believe there is no physical reason against $p_1$ taking negative values. We analyzed both the two-parameter and the three-parameter case. The former presents a strong degeneracy direction in the $a_1$-$p_1$ plane with the presence of long one-sided tails that reduce the constraining power with respect to the HMF. The three-parameter case, on the other hand, provides a robust constraint on all the involved parameters. In our opinion this is the most general fit, with fewer assumptions on the prior information of the parameters, and the one to be used in future works. In fact, it hints at a slightly different behavior of the HMF at intermediate and high masses with respect the traditional ST fit (but still compatible within the uncertainty range).

In this respect, we strongly emphasize that future analyses of the HMF from N-body simulations should provide the range of allowed or explored parameter values used to derived the best-fit because it is an important piece of information. Moreover, based on our conclusions, we would like to recommend the allowance of negative values for $p_1$ in their best-fit calculations.

Lastly, we provided a tabulated form of the constrained $z=0$ HMF for the most representative cases, to be used in future comparisons with updated N-body simulations. As commented in the introduction, these are direct and robust measurements of the HMF.

Concerning further developments, given that this work aims to be a proof of concept, future studies and forthcoming surveys are expected to improve the current constraining power of the submillimeter magnification bias. In this respect, we performed a preliminary analysis, allowing us to draw the following conclusions.

To assess the importance of large-scale uncertainties in the restriction of the HMF, we ran several tests for the three-parameter ST fit using simulated cross-correlation data with smaller error bars, down to an entire order of magnitude. The outcome of such a test was that there are no noticeable changes in the spread of the posterior distributions when significantly reducing the cross-correlation data errors. In addition, the findings of \cite{GON20} about cosmological parameter constraints point in the same direction: increasing the number of sources in an attempt to diminish the statistical errors does not reduce the uncertainties of the results. This is probably related with the use of a single wide redshift bin and the assumption of no time evolution in the astrophysical HOD parameters.

As a consequence, the path forward might lie in performing a tomographic analysis that splits the foreground sample in different bins of redshift. This would likewise allow us to test the suspected time evolution of the HOD parameters as well as that of the HMF parameters. In any case, as for the data error reduction, performing a tomographic analysis will still require an increase in the total number of sources in order to counterbalance the decrease in objects in each bin. On this respect, enlarging both the lenses and the background source samples will increase the statistics per redshift bin in a tomographic analysis.

As for the lenses, the use of the much larger, and already available, sample of sources with optical photometric redshifts might not be straightforward due to their redshift uncertainties. However, currently underway surveys such as DES \citep{DES16} and JPAS \citep{BEN14} might be used in the future for our purposes, given their clear improvement in redshift accuracy. Moreover, the expected \textit{Euclid} mission \citep{LAU11} will certainly provide additional lenses at $z>0.4$.

With respect to the background sources, the already available catalog of the whole area covered by Herschel \citep[HELP,][]{SHI19} can be taken into consideration for the analysis. Moreover, new submillimeter surveys like TolTEC \citep{toltec} or the future mid/near-infrared James Webb Space Telescope \citep[JWST,][]{GAR06} will certainly increase the area and/or the density of the background sources.

\begin{acknowledgements}

MMC, LB and JGN acknowledge the PGC 2018 project PGC2018-101948-B-I00 (MICINN/FEDER). 
AL acknowledges support from PRIN MIUR 2017 prot. 20173ML3WW002, ‘Opening the ALMA window on the cosmic evolution of gas, stars and supermassive black holes’, the MIUR grant ‘Finanziamento annuale individuale attivitá base di ricerca’, and the EU H2020-MSCA-ITN-2019 Project 860744 ‘BiD4BEST: Big Data applications for Black hole Evolution STudies’. 

We deeply acknowledge the CINECA award under the ISCRA initiative, for the availability of high performance computing resources and support. In particular the project “SIS20\_lapi” in the framework “Convenzione triennale SISSA-CINECA”.
In this work, we made extensive use of \texttt{GetDist} \citep{GETDIST}, a Python package for analysing and plotting MC samples. In addition, this research has made use of the python packages \texttt{ipython} \citep{ipython}, \texttt{matplotlib} \citep{matplotlib} and \texttt{Scipy} \citep{scipy}

\end{acknowledgements}

\newpage
\bibliographystyle{aa} % style aa.bst
\bibliography{./main} % your references Yourfile.bib

\appendix

%---------------------------------------
%Appendix A. The ingredients of the model
%---------------------------------------

\section{The ingredients of the model}

The dark matter transfer function has been computed using Eisenstein and Hu's fitting formula \citep{eisenstein98}, which takes baryonic effects into account in a $\Lambda$CDM model. Having chosen an analytical computation of the power spectrum over the traditional numerical one using CAMB \citep{CAMB} is mainly due to computation time. The galaxy-dark matter cross-power spectrum has been computed through equations \eqref{crosspower1h} and \eqref{crosspower2h}, where the linear dark matter power spectrum is evolved to redshift $z$ via the linear growth factor approximation of \cite{carroll92}.

The HMF has of course been parametrized according to \eqref{n_mass} for the two different fits we have described in Section 2.1. The deterministic bias associated with each model has been derived using the peak background split as in \cite{st1999}. 

Furthermore, we have expressed the mean number of galaxies in a halo of mass $M$ as
\begin{equation*}
    \langle N_g\rangle_M=\langle N_{c_g}\rangle_M+\langle N_{s_g}\rangle_M,
\end{equation*}
where $\langle N_{c_g}\rangle_M$ and $\langle N_{s_g}\rangle_M$ are the mean number of central and satellite galaxies in a halo of mass $M$, respectively, expressed in terms of three HOD parameters $(\alpha,\log{M_1},\log{M_{\text{min}}})$ as
\begin{equation*}
    \langle N_{c_g}\rangle_M=\Theta(M-M_{\text{min}})
\end{equation*}
and
\begin{equation*}
    \langle N_{s_g}\rangle_M=\Big(\frac{M}{M_1}\Big)^{\alpha}\Theta(M-M_{\text{min}})
\end{equation*}
following \cite{zehavi05} and \cite{zheng05}. In essence, $M_{\text{min}}$ is the minimum mean halo mass required to host a (central) galaxy and $M_1>M_{\text{min}}$ is the mean halo mass at which exactly one satellite galaxy is hosted. The mean number density of galaxies at redshift $z$ is then given by
\begin{equation*}
    \bar{n}_g(z)=\int_0^{\infty}dM\,n(M,z)\langle N_g\rangle_M.
\end{equation*}

The halo density profile has been assumed to match a Navarro-Frenk-White (NFW) profile \citep{navarro97}. The normalized Fourier transform of the dark matter distribution within a halo of mass $M$ is then given by \citep{cooray02}
\begin{align*}
    u(k,z|M)&=\frac{4\pi\rho_s r_s^3(M,z)}{M}\Big[\sin{kr_s}\big[\text{Si}([1+c]kr_s)-\text{Si}(kr_s)\big]-\\
    &-\frac{\sin{ckr_s}}{[1+c]kr_s}+\cos{kr_s}\big[\text{Ci}([1+c]kr_s)-\text{Ci}(kr_s)\big]\Big],
\end{align*}
where
\begin{equation}
    r_s(M,z)\equiv \frac{R_{\text{vir}}}{c(M,z)}\label{rs}
\end{equation}
and $\rho_s$ are a scale radius and density that parametrize the profile, concentration parameter of a halo of mass $M$ at redshift $z$, which satisfies
\begin{equation}
    M=4\pi \rho_s r_s^3\Big[\ln{[1+c(M,z)]-\frac{c(M,z)}{1+c(M,z)}}\Big]\label{Mnfw}
\end{equation}
for an NFW profile. The virial radius $R_{\text{vir}}$ has been computed through the virial overdensity at redshift $z$, using the fit by \cite{weinberg03}. It should be noted that we have not defined halos at a certain redshift as overdense regions of a constant factor (say 200) times the background or critical density, but using the virial overdensity instead, which depends on redshift. In practice, for a halo of mass $M$, we have adopted the concentration parameter by \cite{bullock01}, computed $r_s$ through \eqref{rs} and, subsequently, calculated $\rho_s$ using \eqref{Mnfw}

%---------------------------------------
%Appendix B. Further discussion results
%---------------------------------------

\onecolumn

\section{Additional tables and figures}

%Table run 5
\begin{table*}[h]
\caption{Parameter priors, marginalized  posterior peaks, means, $68\%$, and $95\%$ credible intervals for run 5 of the MCMC algorithm, that is, a three-parameter ST fit with $0<p_1<0.5$ and fixed HOD values.}
\centering
\begin{tabular}{c c c c c c}
\hline
\hline
Parameter & Prior & Peak & Mean & $68\%$ CI& $95\%$ CI\\
\hline 
$A_1$&$\mathcal{U}$[0,1]&0.59&$0.60$&$[0.47,0.73]$&$[0.35,0.86]$\\
$a_1$&$\mathcal{U}$[0,10]&$0.93$&$1.12$&$[0.65,1.32]$&$[0.40,2.03]$\\
$p_1$&$\mathcal{U}$[0,0.50]&$-$&$-$&$-$&$-$\\
\hline
\hline
\end{tabular}
\label{table_nonnormalization}
\end{table*}

%Corner plot run 5
\begin{figure}[h]
\centering
\includegraphics[width=0.41\columnwidth,height=6.8cm]{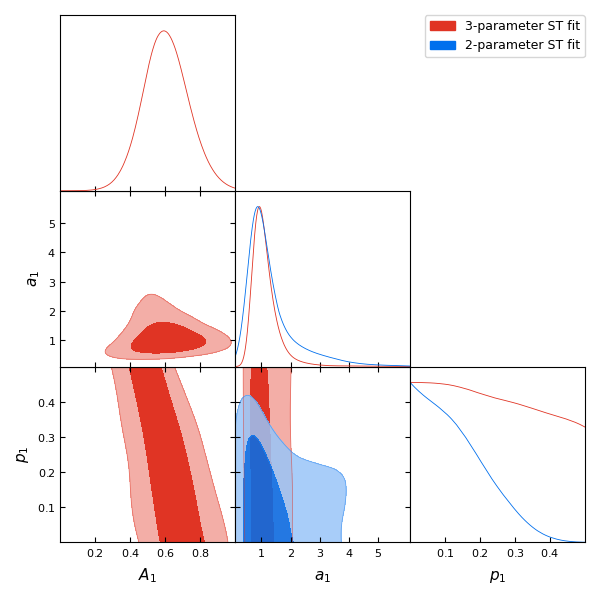}
\caption{One- and two-dimensional (contour) posterior distributions from run 5 (in red) and run 1 (in blue), that is, a three-parameter and a two-parameter ST fit with $0<p_1<0.5$, respectively.}
\label{cornerplot_ST_nonnormalization}
\end{figure}

%Table run 6
\begin{table*}[h]
\caption{Parameter priors, marginalized  posterior peaks, means, $68\%$, and $95\%$ credible intervals for run 6 of the MCMC algorithm, that is, a two-parameter ST fit, $p_1$ allowed to be negative and fixed HOD values.}
\centering
\begin{tabular}{c c c c c c}
\hline
\hline
Parameter & Prior & Peak & Mean & $68\%$ CI& $95\%$ CI\\
\hline 
$a_1$&$\mathcal{U}$[0,10]&$1.46$&$2.86$&$[0.32,3.42]$&$[\,\,\,-\,\,\,,7.78]$\\
$p_1$&$\mathcal{U}$[-10,0.50]&$-0.43$&$-1.24$&$[-1.52,0.31]$&$[-4.36,\,\,\,-\,\,\,]$\\
\hline
\hline
\end{tabular}
\label{table_pneg}
\end{table*}

%Corner plot runs 6 and 7
\begin{figure}[h]
\centering
\includegraphics[width=0.41\columnwidth,height=6.7cm]{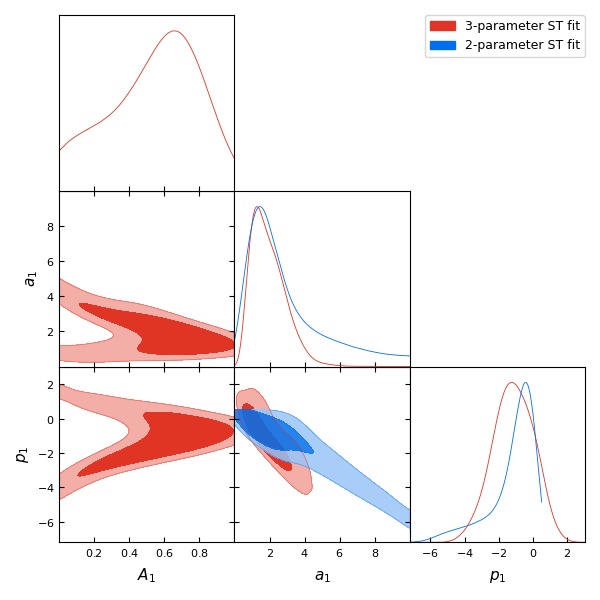}
\caption{One- and two-dimensional (contour) posterior distributions from run 6 (in blue) and run 7 (in red), that is, a two-parameter ST fit with $-10<p_1<0.5$ and fixed HOD values and a three-parameter ST fit with $-10<p_1<10$ and fixed HOD values, respectively.}
\label{cornerplot_ST_pneg}
\end{figure}

%Corner plot runs 1 and 2
\begin{figure*}[h]
\centering
\includegraphics[width=0.8\textwidth,height=14.3cm]{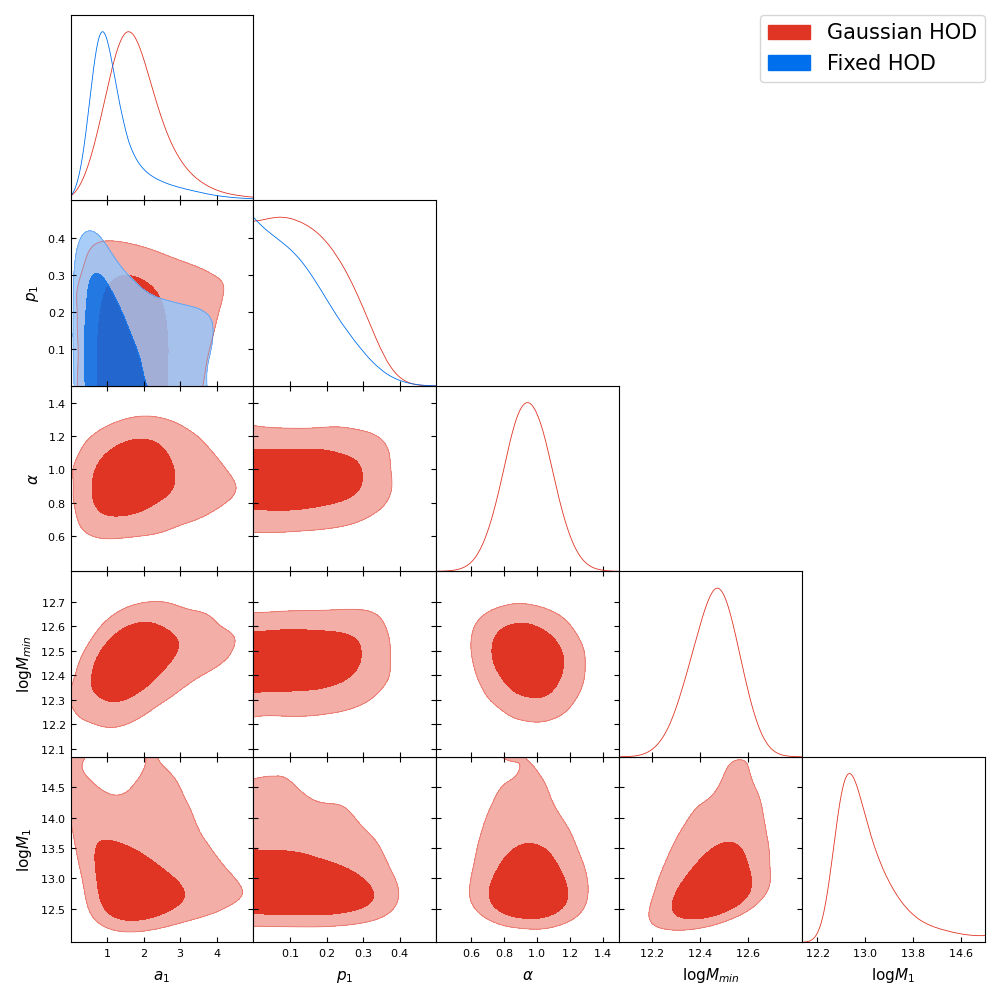}
\caption{One- and two-dimensional (contour) posterior distributions from run 1 (in blue) and run 2 (in red), that is, a two-parameter ST fit with fixed values and with Gaussian priors for the HOD parameters, respectively. The $p_1$ parameter is assumed to be positive.}
\label{cornerplot_ST_fixedvsgaussian}
\end{figure*}

%ST (fixed and gaussian)
\begin{table*}[t]
\caption{Tabulation of the $z=0$ HMF at as obtained via the sampling of the full posterior for the two-parameter ST fit in the two cases studied in Section 4.1. For convenience, we have tabulated the base-10 logarithm of all quantities; the masses are expressed in $M_{\odot}/h$ and the median, lower and upper bounds of the credible intervals are expressed in $h^4\,\text{Mpc}^{-3}M_{\odot}
^{-1}$. }
\centering
\begin{tabular}{c| c c c| c c c}
\hline
\hline
 & \multicolumn{3}{|c|}{two-parameter ST fit: fixed HOD} & \multicolumn{3}{c}{two-parameter ST fit: Gaussian HOD} \\
$\log_{10}{M}$& $\log_{10}{\text{Med}}$ & $\log_{10}{68\% \,\text{CI}}$&$\log_{10}{95\%\, \text{CI}}$ &$\log_{10}{\text{Med}}$ & $\log_{10}{68\% \,\text{CI}}$&$\log_{10}{95\%\, \text{CI}}$\\
\hline 
10.0&-10.54&$[-10.62,-10.47]$&$[-10.66,-10.38]$ &-10.48&$[-10.53,-10.41]$&$[-10.65,-10.37]$\\
10.5&-11.47&$[-11.54,-11.40]$&$[-11.60,-11.33]$&-11.42&$[-11.47,-11.35]$&$[-11.60,-11.33]$\\
11.0&-12.40&$[-12.46,-12.33]$&$[-12.54,-12.29]$&-12.37&$[-12.41,-12.30]$&$[-12.55,-12.29]$\\
11.5&-13.34&$[-13.38,-13.27]$&$[-13.49,-13.25]$&-13.33&$[-13.37,-13.25]$&$[-13.52,-13.25]$\\
12.0&-14.29&$[-14.32,-14.22]$&$[-14.46,-14.22]$&-14.30&$[-14.35,-14.22]$&$[-14.53,-14.22]$\\
12.5&-15.24&$[-15.28,-15.18]$&$[-15.49,-15.18]$&-15.30&$[-15.36,-15.18]$&$[-15.60,-15.18]$\\
13.0&-16.23&$[-16.29,-16.14]$&$[-16.65,-16.14]$&-16.35&$[-16.46,-16.17]$&$[-16.80,-16.14]$\\
13.5&-17.26&$[-17.36,-17.11]$&$[-18.02,-17.10]$&-17.49&$[-17.67,-17.15]$&$[-18.23,-17.10]$\\
14.0&-18.39&$[-18.57,-18.12]$&$[-19.78,-18.10]$&-18.83&$[-19.20,-18.29]$&$[-20.06,-18.05]$\\
14.5&-19.72&$[-20.06,-19.25]$&$[-22.26,-19.19]$&-20.54&$[-21.23,-19.59]$&$[-22.73,-19.05]$\\
15.0&-21.51&$[-22.18,-20.59]$&$[-26.26,-20.49]$&-23.05&$[-24.32,-21.23]$&$[-27.04,-20.07]$\\
15.5&-24.34&$[-25.73,-22.48]$&$[-33.69,-22.22]$&-27.38&$[-29.88,-23.74]$&$[-35.11,-21.25]$\\
\hline
\hline
\end{tabular}
\label{tabulationST}
\end{table*}

%Corner plot runs 3 and 4
\begin{figure*}[hbtp]
\centering
\includegraphics[width=0.8\textwidth]{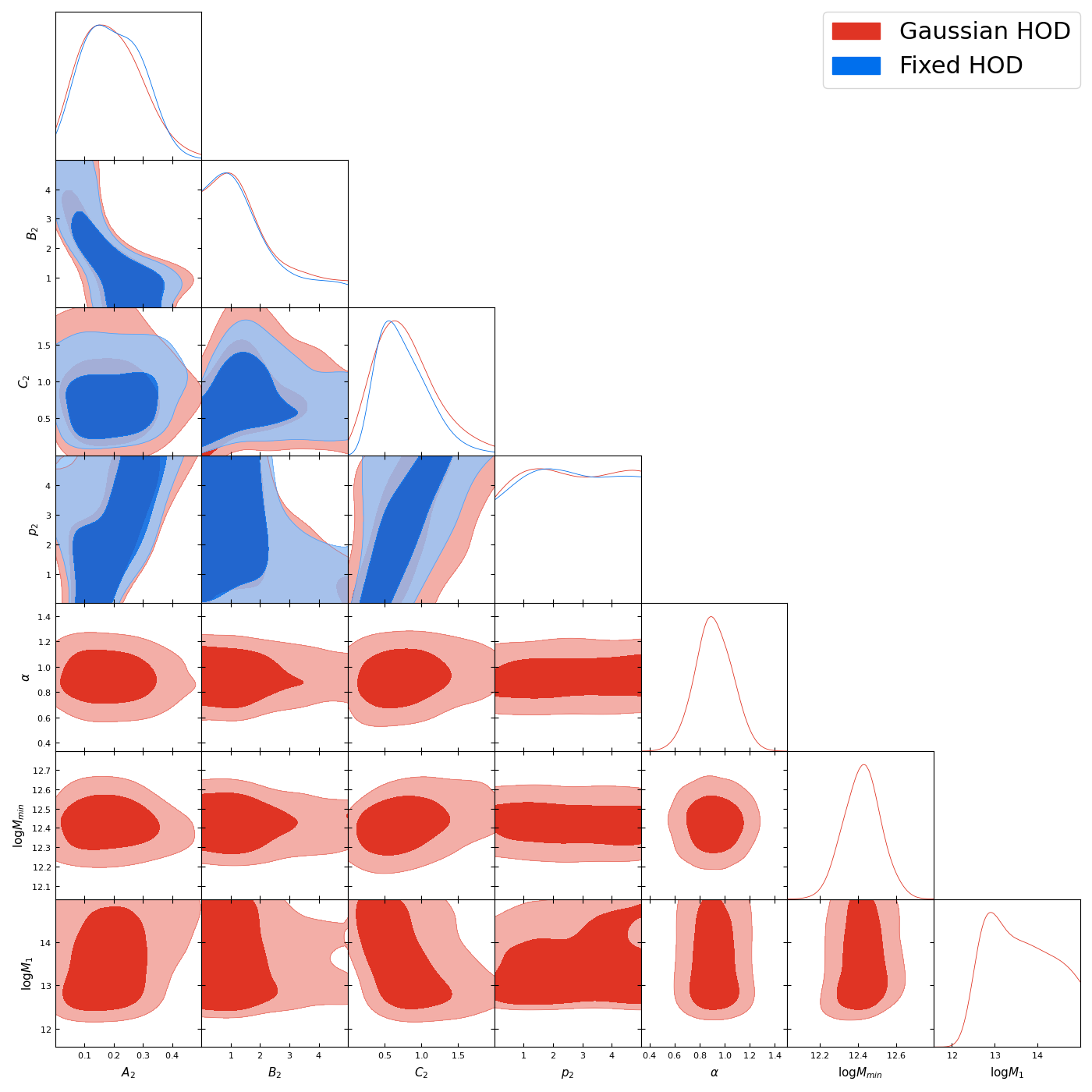}
\caption{One- and two-dimensional (contour) posterior distributions from run 3 (in blue) and run 4 (in red), that is, a four-parameter Tinker fit with fixed values and with Gaussian priors for the HOD parameters, respectively. The $p_2$ parameter is assumed to be positive.}
\label{cornerplot_tinker_fixedvsgaussian}
\end{figure*}

%Tinker (fixed and gaussian)
\begin{table*}[t]
\caption{Tabulation of the $z=0$ HMF at as obtained via the sampling of the full posterior for the four-parameter Tinker fit in the two cases studied in Section 4.2. For convenience, we have tabulated the base-10 logarithm of all quantities; the masses are expressed in $M_{\odot}/h$ and the median, lower and upper bounds of the credible intervals are expressed in $h^4\,\text{Mpc}^{-3}M_{\odot}
^{-1}$. }
\centering
\begin{tabular}{c| c c c| c c c}
\hline
\hline
 & \multicolumn{3}{|c|}{four-parameter Tinker fit: fixed HOD} & \multicolumn{3}{c}{four-parameter Tinker fit: Gaussian HOD} \\
$\log_{10}{M}$& $\log_{10}{\text{Med}}$ & $\log_{10}{68\% \,\text{CI}}$&$\log_{10}{95\%\, \text{CI}}$ &$\log_{10}{\text{Med}}$ & $\log_{10}{68\% \,\text{CI}}$&$\log_{10}{95\%\, \text{CI}}$\\
\hline 
10.0&-10.36&$[-10.45,-10.25]$&$[-10.68,-10.18]$ &-10.39&$[-10.51,-10.24]$&$[-10.72,-10.13]$\\
10.5&-11.32&$[-11.39,-11.21]$&$[-11.58,-11.15]$&-11.35&$[-11.48,-11.21]$&$[-11.64,-11.09]$\\
11.0&-12.28&$[-12.35,-12.18]$&$[-12.49,-12.10]$&-12.30&$[-12.42,-12.15]$&$[-12.59,-12.04]$\\
11.5&-13.23&$[-13.32,-13.12]$&$[-13.46,-13.03]$&-13.24&$[-13.39,-13.08]$&$[-13.58,-12.93]$\\
12.0&-14.18&$[-14.33,-14.04]$&$[-14.45,-13.89]$&-14.20&$[-14.40,-14.00]$&$[-14.60,-13.75]$\\
12.5&-15.16&$[-15.35,-14.96]$&$[-15.50,-14.73]$&-15.17&$[-15.45,-14.91]$&$[-15.69,-14.57]$\\
13.0&-16.15&$[-16.41,-15.90]$&$[-16.58,-15.57]$&-16.17&$[-16.51,-15.82]$&$[-16.86,-15.38]$\\
13.5&-17.20&$[-17.50,-16.88]$&$[-17.75,-16.45]$&-17.23&$[-17.68,-16.79]$&$[-18.21,-16.20]$\\
14.0&-18.36&$[-18.68,-17.95]$&$[-19.19,-17.53]$&-18.40&$[-19.01,-17.83]$&$[-19.85,-17.08]$\\
14.5&-19.74&$[-20.06,-19.23]$&$[-21.06,-18.81]$&-19.86&$[-20.61,-18.90]$&$[-22.13,-18.16]$\\
15.0&-21.73&$[-22.32,-20.85]$&$[-24.38,-20.56]$&-21.98&$[-22.97,-20.10]$&$[-26.07,-19.62]$\\
15.5&-25.22&$[-26.58,-22.94]$&$[-30.68,-22.28]$&-25.67&$[-28.09,-22.27]$&$[-33.76,-21.16]$\\
\hline
\hline
\end{tabular}
\label{tabulationtinker}
\end{table*}

%3-parameter ST "free"
\begin{table*}[t]
\caption{Tabulation of the $z=0$ HMF at as obtained via the sampling of the full posterior for the three-parameter ST fit with $-10<p_1<10$ and fixed HOD. For convenience, we have tabulated the base-10 logarithm of all quantities; the masses are expressed in $M_{\odot}/h$ and the median, lower and upper bounds of the credible intervals are expressed in $h^4\,\text{Mpc}^{-3}M_{\odot}
^{-1}$. }
\centering
\begin{tabular}{c| c c c}
\hline
\hline
$\log_{10}{M}$& $\log_{10}{\text{Med}}$ & $\log_{10}{68\% \,\text{CI}}$&$\log_{10}{95\%\, \text{CI}}$\\
\hline 
10.0&-10.50&$[-10.62,-10.31]$&$[-11.12,-10.22]$\\
10.5&-11.41&$[-11.52,-11.23]$&$[-11.97,-11.18]$\\
11.0&-12.31&$[-12.42,-12.15]$&$[-12.79,-12.10]$\\
11.5&-13.20&$[-13.29,-13.06]$&$[-13.54,-13.00]$\\
12.0&-14.05&$[-14.15,-13.92]$&$[-14.48,-13.84]$\\
12.5&-14.92&$[-15.09,-14.69]$&$[-15.51,-14.50]$\\
13.0&-15.84&$[-16.09,-15.48]$&$[-16.60,-15.14]$\\
13.5&-16.81&$[-17.16,-16.38]$&$[-17.73,-15.89]$\\
14.0&-17.93&$[-18.34,-17.46]$&$[-19.07,-16.96]$\\
14.5&-19.37&$[-19.76,-18.85]$&$[-20.85,-18.49]$\\
15.0&-21.62&$[-22.24,-20.70]$&$[-23.94,-20.39]$\\
15.5&-25.72&$[-27.38,-23.01]$&$[-31.15,-22.40]$\\
\hline
\hline
\end{tabular}
\label{tabulationfree}
\end{table*}

%%%% End of aa.dem
\end{document}